\documentclass[sigconf]{acmart}

\AtBeginDocument{%
  }

\setcopyright{acmlicensed}
\copyrightyear{2018}
\acmYear{2018}
\acmDOI{XXXXXXX.XXXXXXX}
\acmConference[Conference acronym 'XX]{Make sure to enter the correct
  conference title from your rights confirmation email}{June 03--05,
  2018}{Woodstock, NY}
\acmISBN{978-1-4503-XXXX-X/2018/06}

\PassOptionsToPackage{dvipsnames,table}{xcolor}
\usepackage{xcolor}
\usepackage{booktabs} % For formal tables
\usepackage{tabularx}
\usepackage{colortbl}
\usepackage{arydshln}
\usepackage{adjustbox}
\usepackage[linesnumbered,ruled,vlined]{algorithm2e}
\usepackage{amsmath}
 
\usepackage{amssymb}
\usepackage{bm}
\usepackage{enumitem}
\usepackage{soul}
\usepackage{threeparttable}  
\usepackage{multirow}
\usepackage{makecell}
\usepackage{tabularx}
\usepackage[T1]{fontenc}
\usepackage{array}
\usepackage{listings}
\usepackage{xspace} 
\usepackage{subcaption}
\usepackage{graphicx}
\usepackage{hyperref}
\usepackage{textcomp}
\usepackage{microtype} 
\usepackage{calc}
\usepackage{pifont}
\usepackage{tcolorbox}
\usepackage{lscape}
\usepackage{rotating}
\usepackage{tikz}
\usepackage{mdframed}
\usetikzlibrary{positioning}
\usepackage{balance}
\usepackage{amsfonts}

\usepackage{wrapfig} 

\newcommand{\distance}{5pt}
\newcolumntype{Y}{>{\centering\arraybackslash}X}
\newcolumntype{Z}{>{\raggedright\arraybackslash}X}

\SetCommentSty{mycommfont}
\SetKwInput{KwInput}{Input}                % Set the Input
\SetKwInput{KwOutput}{Output}              % set the Output

\newlength{\ColorBoxDepthReference}
\newlength{\ColorBoxHeightReference}
\newlength{\Width}%
\newcommand{\MyColorBox}[2][red]%
{%
	\settowidth{\Width}{#2}%
	\colorbox{#1}%
	{%      
		\raisebox{-\ColorBoxDepthReference}%
		{%
			\parbox[b][\ColorBoxHeightReference+\ColorBoxDepthReference][c]{\Width}{\centering#2}%
		}%
	}%
}
\definecolor{codegreen}{rgb}{0,0.6,0}
\definecolor{codegray}{rgb}{0.5,0.5,0.5}
\definecolor{codepurple}{rgb}{0.58,0,0.82}
\definecolor{border_gray}{RGB}{80, 80, 80}    % 深灰边框
\definecolor{header_gray}{RGB}{240, 240, 240} % 描述区浅灰
\definecolor{patch_gray}{RGB}{250, 250, 250}  % Agent Patch 极浅灰
\definecolor{gold_white}{RGB}{255, 255, 255}  % Gold Patch 纯白
\newcommand{\todo}[1]{\textcolor{red}{TODO: #1}}

\lstdefinestyle{minicode}{
  basicstyle=\ttfamily\fontsize{6pt}{6.5pt}\selectfont,
  columns=fullflexible,
  breaklines=true,
  showstringspaces=false,
  aboveskip=0pt,
  belowskip=0pt,
  keywordstyle=\bfseries, 
  commentstyle=\color{gray},
  rulecolor=\color{black!10}
}

\newtcolorbox{patchbox}[3][]{
  colback=#2,
  colframe=black!15, 
  boxrule=0.4pt,
  sharp corners,
  title={#3},
  fonttitle=\bfseries\fontsize{5.5pt}{6pt}\selectfont,
  coltitle=black,
  left=2pt, 
  right=2pt, 
  top=1pt,            % 不要用太大的负值，改为较小的正值
  bottom=1pt,         % 显著减小底部边距
  boxsep=1pt,         % 减小全局边距
  middle=0pt,         % 标题与内容之间的间距设为 0
  nobeforeafter,
  width=\linewidth
}

\newtcolorbox{compactcase}[1]{
  colback=white,
  colframe=border_gray,
  boxrule=0.6pt,
  sharp corners,
  left=0pt, right=0pt, top=0pt, bottom=1pt,
  boxsep=0pt,
  fonttitle=\bfseries\scriptsize\color{black},
  colbacktitle=header_gray, 
  title={#1},
  before skip=6pt, after skip=6pt,
  toptitle=1.5pt, bottomtitle=1.5pt
}

\newcommand{\eg}{\textit{e.g.,}\xspace}
\newcommand{\ie}{\textit{i.e.,}\xspace}

\newcommand{\bench}{\textsc{SWE-Shield}\xspace}
\newcommand{\benchVerified}{\textsc{SWE-Shield}$_{verified}$\xspace}
\newcommand{\benchPro}{\textsc{SWE-Shield}$_{pro}$\xspace}
\newcommand{\tool}{\textsc{DesignHunter}\xspace}

\newcommand{\chong}[1]{\textcolor{blue}{\textbf{Chong:} #1}}

\begin{document}


\title[Does Pass Rate Tell the Whole Story? Evaluating Design Constraint Compliance in LLM-based Issue Resolution]{Does Pass Rate Tell the Whole Story? Evaluating Design Constraint Compliance in LLM-based Issue Resolution}

% \title[\bench: Beyond Passing Tests to Mastering Design Constraints in LLM-based Issue Resolution]{\bench: Beyond Passing Tests to Mastering Design Constraints in LLM-based Issue Resolution}

% \affiliation{%
\author{Kai Yu}
\email{23262010042@m.fudan.edu.cn}
\affiliation{%
  \institution{Fudan University}
  \country{China}
}

\author{Zhenhao Zhou}
\email{zhzhou24@m.fudan.edu.cn}
\affiliation{%
  \institution{Fudan University}
  \country{China}
}

\author{Junhao Zeng}
\email{alanzeng423@outlook.com}
\affiliation{%
  \institution{Fudan University}
  \country{China}
}

\author{Ying Wang}
\email{yingwang25@m.fudan.edu.cn}
\affiliation{%
  \institution{Fudan University}
  \country{China}
}

\author{Xueying Du}
\email{21210240012@m.fudan.edu.cn}
\affiliation{%
  \institution{Fudan University}
  \country{China}
}

\author{Zhiqiang Yuan}
\email{zhiqiangyuan23@m.fudan.edu.cn}
\affiliation{%
  \institution{Fudan University}
  \country{China}
}

\author{Junwei Liu}
\email{jwliu24@m.fudan.edu.cn}
\affiliation{%
  \institution{Fudan University}
  \country{China}
}

\author{Ziyu Zhou}
\email{25113050202@m.fudan.edu.cn}
\affiliation{%
  \institution{Fudan University}
  \country{China}
}

\author{Yujia Wang}
\email{yujiawang24@m.fudan.edu.cn}
\affiliation{%
  \institution{Fudan University}
  \country{China}
}

\author{Chong Wang}
\authornote{Chong Wang and Xin Peng are corresponding authors.}
\email{chong.wang@ntu.edu.sg}
\affiliation{%
  \institution{Nanyang Technological University}
  \country{Singapore}
}

\author{Xin Peng}
\authornotemark[1]
\email{pengxin@fudan.edu.cn}
\affiliation{%
  \institution{Fudan University}
  \country{China}
}

% \affiliation{%

% \affiliation{%

\renewcommand{\shortauthors}{Yu et al.}

\begin{abstract}

Repository-level issue resolution benchmarks have become a standard testbed for evaluating LLM-based agents, yet success is still predominantly measured by test pass rates. In practice, however, acceptable patches must also comply with project-specific design constraints, such as architectural conventions, error-handling policies, and maintainability requirements, which are rarely encoded in tests and are often documented only implicitly in code review discussions. This paper introduces \textit{design-aware issue resolution} and presents \bench{}, a benchmark that makes such implicit design constraints explicit and measurable. \bench{} is constructed by mining and validating design constraints from real-world pull requests, linking them to issue instances, and automatically checking patch compliance using an LLM-based verifier, yielding 495 issues and 1,787 validated constraints across six repositories, aligned with SWE-bench-Verified and SWE-bench-Pro. Experiments with state-of-the-art agents show that test-based correctness substantially overestimates patch quality: fewer than half of resolved issues are fully design-satisfying, design violations are widespread, and functional correctness exhibits negligible statistical association with design satisfaction. While providing issue-specific design guidance reduces violations, substantial non-compliance remains, highlighting a fundamental gap in current agent capabilities and motivating design-aware evaluation beyond functional correctness.

\end{abstract}


\keywords{Issue Resolution, Large Language Model, Design Constraint}
\begin{CCSXML}
<ccs2012>
   <concept>
       <concept_id>10011007</concept_id>
       <concept_desc>Software and its engineering</concept_desc>
       <concept_significance>500</concept_significance>
       </concept>
 </ccs2012>
\end{CCSXML}

\ccsdesc[500]{Software and its engineering}

\newcommand{\summary}[1]{
 \begin{center}
  \begin{tcolorbox}[colback=gray!10,colframe=black!25,width=1\columnwidth,arc=1mm, auto outer arc,boxrule=0.5pt,boxsep=2pt,left=3pt,right=3pt,top=0pt,bottom=0pt]
  \textbf{SUMMARY:} #1
  \end{tcolorbox}
 \end{center}
}

\newcommand{\key}[1]{
 \begin{center}
  \begin{tcolorbox}[colback=gray!10,colframe=black!25,width=1\columnwidth,arc=1mm, auto outer arc,boxrule=0.5pt,boxsep=2pt,left=3pt,right=3pt,top=0pt,bottom=0pt]
  \textbf{} #1
  \end{tcolorbox}
 \end{center}
}

\maketitle
\section{Introduction}
Large language models (LLMs) and LLM-based agents show strong potential in software engineering tasks such as code generation~\cite{DBLP:journals/corr/abs-2107-03374, DBLP:journals/corr/abs-2406-00515, DBLP:journals/corr/abs-2508-00083}, defect detection~\cite{zhou2019devign, lu2021codexglue}, and code summarization~\cite{sun2024source, ahmed2022few}. To better assess their effectiveness in realistic software engineering settings, recent research has increasingly focused on real-world issue resolution, a core activity in software maintenance. Consequently, several benchmarks, such as SWE-bench~\cite{DBLP:conf/iclr/JimenezYWYPPN24}, have been proposed, fueling intense leaderboard-driven evaluation. These benchmarks provide an initial snapshot of the capabilities and limitations of LLMs and LLM-based agents in real-world software development scenarios.

As an early effort in this direction, SWE-bench~\cite{DBLP:conf/iclr/JimenezYWYPPN24} collects real-world issues from highly starred open-source projects (\eg Django~\cite{django_github}) on GitHub. However, most of the collected projects are \textit{libraries or frameworks}, which primarily aim to provide diverse APIs rather than implement complex business logic or intricate functional interactions. To extend benchmarking toward more complex scenarios, subsequent benchmarks such as SWE-bench Pro~\cite{DBLP:journals/corr/abs-2509-16941} and SWE-Lancer~\cite{DBLP:conf/icml/MiserendinoWPH25} focus on \textit{enterprise-level software}, where issue resolution involves longer horizons and greater difficulty. Complementing these efforts, SWE-bench Multimodal~\cite{DBLP:conf/iclr/YangJZLYWPMSNY025} augments the original benchmark with issues that include \textit{visual elements} (\eg bug screenshots), thereby evaluating models' ability to interpret and act on information presented across both textual and visual modalities. The evolution of these benchmarks reflects a clear shift toward evaluating LLM/agent capabilities in more realistic enterprise software maintenance practices.

Although the issue complexity in these benchmarks is more closely aligned with enterprise practices, existing evaluations of resolution effectiveness primarily focus on \textit{functional correctness}, typically measured by the pass rate on test cases. Specifically, a generated patch is deemed to have successfully resolved the issue if it passes all predefined tests. However, in real-world software development, patch acceptance depends on more than test outcomes; \textit{it is also governed by multidimensional \textbf{design constraints}, ranging from project-wide conventions to scenario-specific trade-offs.} Following ISO-29148~\cite{ISO29148_2011}, we define design constraints as requirements that restrict a designer's options by imposing immovable boundaries and limits. For instance, a well-maintained project may enforce a convention that certain methods should propagate exceptions to higher-level callers rather than catching and handling them locally. Violating such constraints can lead to a patch, whether produced by a human developer or an LLM or agent, being rejected even when all test cases pass.

To bridge this gap, we propose evaluating LLMs and LLM-based agents on their awareness of and compliance with design constraints in issue resolution, offering a critical perspective that goes beyond simple pass rates. Achieving such a design-aware evaluation presents two main challenges. \textit{First, design constraints are rarely documented explicitly and are instead embedded implicitly in a project's evolution history.} In GitHub projects, such constraints are often expressed through pull requests, including associated code reviews and discussion threads. However, multiple design constraints may be intertwined within a single review comment or discussion thread, while a single constraint may be distributed across multiple pull requests and review discussions, each capturing only a partial specification of the constraint. \textit{Second, applying these constraints and validating compliance to construct a design-aware benchmark is non-trivial.} For each issue, it is necessary to identify relevant design constraints and provide a method for automatic validation. This involves linking issues to constraints, possibly extracted from pull requests corresponding to other issues, such as a convention requiring the use of \texttt{f}-strings for string formatting. In addition, the benchmark must determine whether a constraint is satisfied based on the reasoning traces and patches generated by LLMs or LLM-based agents.

We construct a novel benchmark, \textbf{\bench}, to evaluate the effectiveness of LLMs and LLM-based agents in \textit{design-aware issue resolution}. \bench is built through a multi-stage pipeline that distills implicit design knowledge from real-world software development artifacts and integrates it into issue resolution tasks. Starting from pull requests in large-scale code repositories, we automatically extract design constraints using \tool, a newly proposed LLM-based two-stage extraction approach. These constraints capture generalized yet context-aware design guidance grounded in developer-authored code review discussions. The extracted constraints are then associated with issues resolved by merged pull requests through a combination of explicit traceability and semantic matching, followed by targeted manual validation to ensure fidelity. To support evaluation beyond test-based correctness, \bench further incorporates an LLM-based patch verifier that assesses whether generated patches satisfy the associated design constraints. The resulting benchmark includes two variants, \benchVerified and \benchPro, derived from SWE-bench-Verified~\cite{DBLP:conf/iclr/JimenezYWYPPN24} and SWE-bench Pro~\cite{DBLP:journals/corr/abs-2509-16941}, respectively, and comprises hundreds of real-world issues and thousands of manually verified design constraints explicitly linked to historical code review evidence.

We conduct extensive experiments on \bench{} and obtain four key findings.
First, agents achieve high functional correctness but limited design compliance: Pass Rate reaches 70.25\%--75.95\% on \benchVerified{} and up to 42.69\% on \benchPro{}, while design satisfaction remains low (DSR=32.64\%--50.20\%) and violations are common (DVR up to 45.85\%).
Second, functional correctness is a poor proxy for design compliance: a $\chi^2$ test shows no significant association in most settings, with consistently negligible effect sizes (Cramér’s $V \le 0.11$), and many test-passing patches still violate applicable design constraints.
Third, model choice yields only modest improvements in design satisfaction despite large gaps in Pass Rate: under the same \textsc{swe-agent} framework on \benchPro{}, DSR varies within 12 percentage points across foundation models, and violated-constraint analysis reveals a substantial shared core missed by all models.
Finally, providing issue-specific design-constraint guidance reduces violations (DVR decreases by up to 6.35 percentage points), but residual violation rates remain above 30\%.
Overall, these results show that test-based metrics substantially overestimate patch quality and underscore the need for explicit design-aware evaluation beyond functional correctness.

In summary, this paper makes the following main contributions:
\begin{itemize}[leftmargin=15pt]
    \item \textbf{\bench}, a novel benchmark that enables the evaluation of LLM-based issue resolution with respect to both functional correctness and compliance with design constraints. \bench consists of 495 issues from 6 projects, associated with 1,787 design constraints.

    \item \textbf{\tool}, an LLM-based approach for extracting design constraints from pull requests. Using \tool, we identify a total of 10,885 design constraints.
    
    \item \textbf{Extensive empirical results}, which reveals that state-of-the-art LLMs and agents still face significant challenges in meeting design constraints.
\end{itemize}

\section{Motivation}
\begin{figure*}[tbp]
    \centering
    \includegraphics[width=\textwidth]{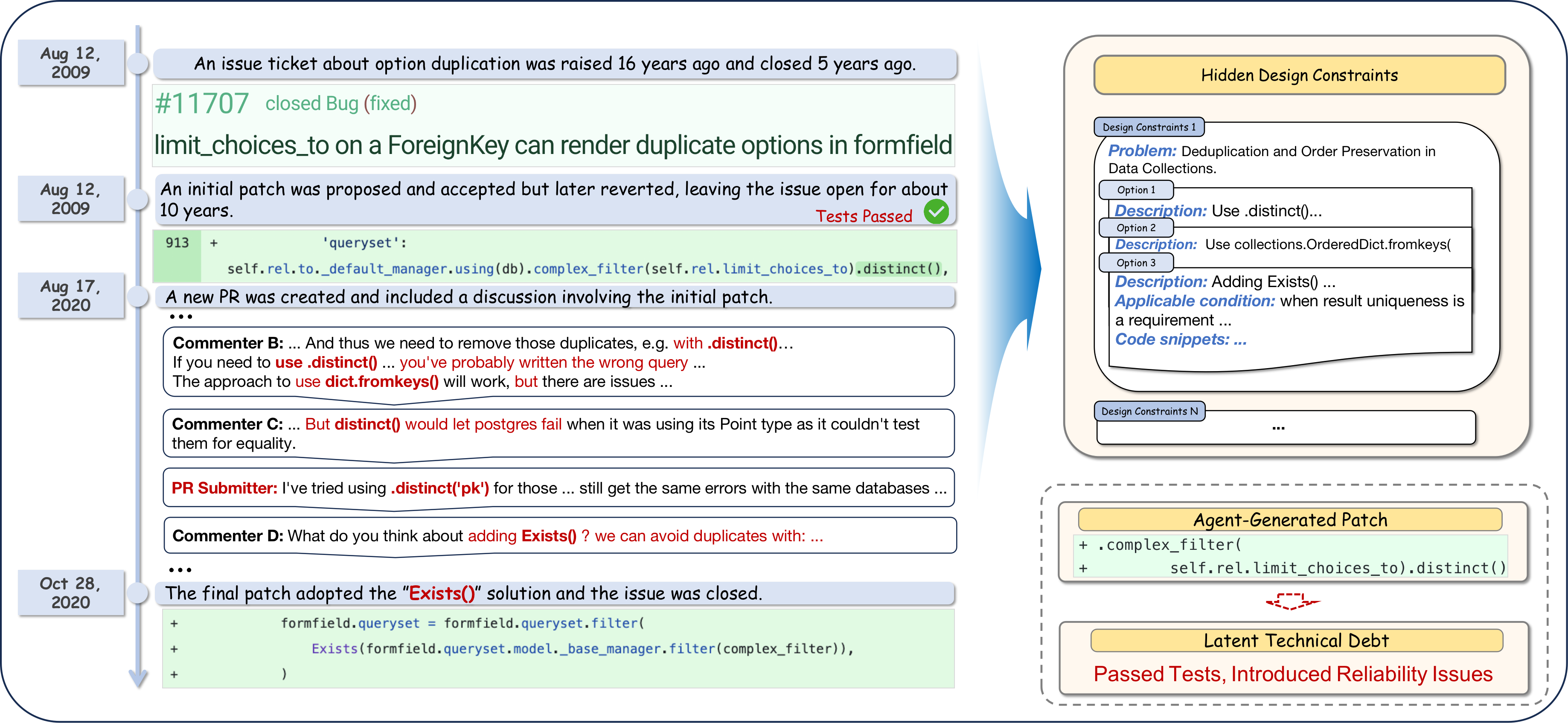}
    \caption{A motivating example illustrating the resolution process of a realistic issue, along with relevant code review threads and the design constraints embedded in the discussion.}
    \label{fig:motivation}
\end{figure*}

We present a motivating example to illustrate why design constraints must be considered during issue resolution, and to highlight the challenges of acquiring and verifying such constraints given their largely implicit nature. Figure~\ref{fig:motivation} presents the example derived from the timeline of a real-world issue resolution in the Django project~\cite{django_github}. Although the original resolution spans more than ten years and involves complex discussions, we simplify it here to better illustrate our core motivation.

\subsection{The Necessity of Mastering Design Constraints in Issue Resolution}
In practice, whether a patch is accepted during issue resolution depends not only on its functional correctness, as indicated by passing test cases, but also on whether it complies with the design constraints associated with the issue and the affected code context.

The issue shown in Figure~\ref{fig:motivation} concerns a duplication problem when interacting with databases. An initial patch was promptly proposed by adding a simple \texttt{distinct()} call to the original code logic. This patch was quickly accepted because it passed all test cases. However, it was later reverted through another ticket after developers observed that \texttt{distinct()} failed to handle duplication correctly when PostgreSQL was used. As a result, the issue remained open for nearly ten years until a new pull request was submitted.
In this pull request, a code review thread involving multiple developers revisited the problem and discussed three alternative solutions: \texttt{distinct()}, \texttt{dict.fromkeys()}, and \texttt{Exists()}. The reviewers analyzed the advantages, limitations, and applicable scenarios of each option, as well as their potential consequences. Ultimately, the \texttt{Exists()} solution was adopted, and the issue was finally closed.

This case highlights several design constraints that go beyond the explicit functional requirements of the issue. Although the problem description itself was clear, many important design considerations were implicit and external to the issue report. As Django is a widely used web framework, applications may connect to diverse database systems depending on deployment environments. Therefore, resolving such issues requires awareness of latent external dependencies and potential reliability concerns, rather than relying solely on existing test cases. Ignoring these constraints risks introducing long-term technical debt.

\subsection{The Pitfall of Evaluating AI Assistants Solely Based on Test Cases}
This issue is included in the widely used issue-resolution benchmark SWE-bench~\cite{DBLP:conf/iclr/JimenezYWYPPN24}. We applied a state-of-the-art agent tool, Live-SWE-agent, powered by one of the most advanced LLMs (Gemini~3~Pro), to generate a patch for this issue. The agent also employed \texttt{distinct()} and passed all test cases provided by SWE-bench, thereby being deemed successful under the benchmark's evaluation criteria.
However, as demonstrated by the earlier analysis of the code review discussions, this patch would not be accepted in real-world development. This discrepancy suggests that current leaderboards, which rely primarily on test-case outcomes, fail to comprehensively reflect the practical usability of AI-assisted issue resolution tools.

One might argue that expanding test coverage could address this limitation by exposing additional failure cases. While increased coverage is beneficial, \textit{we contend that evaluation should also account for higher-level design considerations}. Our review of multiple pull request discussions reveals recurring types of design constraints that significantly influence patch acceptance.\footnote{In this work, we do not attempt to systematically categorize design constraints; we leave this as an important direction for future research.} These constraints arise at various scopes, including project-level conventions (\eg exception handling patterns, logging practices, and the use of \texttt{f}-strings), context-dependent design decisions, scenario-specific trade-offs among correctness, performance, and maintainability, cross-cutting concerns such as functionality reuse and API consistency, and expectations regarding implementation style and code organization.
During decision-making, many of the design factors discussed above must be taken into account, yet they are often difficult to capture through test execution alone. 

Together, these observations motivate the need for a design-aware evaluation perspective for LLM-based issue resolution, one that goes beyond test-based correctness and considers alignment with the design decisions that have emerged throughout a project’s evolution.

\subsection{The Challenges of Acquiring and Verifying Implicit Design Constraints}

\textbf{The Implicit Nature of Design Knowledge.}
Even when design knowledge appears in development artifacts such as pull requests, it is rarely expressed in an isolated or well-structured form. In practice, multiple design considerations are often \emph{entangled within a single review comment}, while the same concern may be \emph{scattered across different comments or pull requests}. A single comment can address performance, modularity, and API consistency simultaneously, and a solution may be proposed without explicit justification, with its rationale documented elsewhere.
This entanglement and dispersion make it difficult to extract coherent and reusable design constraints from raw discussions, motivating the need for fine-grained techniques that jointly consider structural and semantic information in project histories.

% Even when design knowledge is present in development artifacts such as pull requests, it is rarely expressed as isolated or well-structured information. In practice, multiple design considerations are often \emph{entangled within a single review comment}, while the same design concern may be \emph{scattered across different comments or pull requests}. A single comment may simultaneously address performance, modularity, and API consistency, thereby intertwining multiple design decisions. Moreover, a solution proposal may appear without explicit justification in one pull request, while its underlying rationale is documented elsewhere.
% This inherent entanglement and dispersion make it difficult to simply extract coherent and reusable design constraints from raw textual discussions. Consequently, more fine-grained techniques are needed to mine design constraints from project histories by jointly considering their structural and semantic characteristics.

\textbf{The Lack of Reliable Verification Methods.}
Unlike test cases, design constraints are \emph{non-executable} and \emph{context-dependent}, regardless of whether they are represented in structured or unstructured forms. The same design decision may be realized through different code structures depending on contextual factors. For instance, a constraint related to asynchronization can be implemented in multiple ways, making it difficult to verify through predefined ground-truth.
As a result, design-aware evaluation cannot rely on traditional oracle-based verification. Instead, it requires semantic comparison between code implementations and design constraints to determine whether a patch aligns with the intended design requirements.

% \subsection{Approach Comparison and Our Technique}

% Existing benchmarks and datasets for issue resolution primarily expand along the breadth dimension—by increasing task diversity, programming languages, or repository scale—while largely treating any test-passing patch as acceptable...

% At the same time, prior work on design decision extraction often directly extracts design decisions from development artifacts without resolving the problem of mixing and scattering, and without grounding the extracted decisions in code, resulting in fragmented knowledge that is difficult to apply to patch evaluation.

% \textbf{Our Design-Aware Idea.}
% We introduce design alignment as an orthogonal evaluation dimension for issue resolution, and ground it in code by extracting atomic, code-linked design claims from review discussions and aggregating them across contexts to reconstruct coherent design constraints that can be used to assess whether a patch is not only correct, but also design-consistent.

%We define Design Alignment Rate (DAR) as a quantitative measure to assess the degree to which an LLM-generated patch adheres to the underlying design decisions and quality-attribute trade-offs of a software system.
\section{Construction of \bench}

\begin{figure*}
    \centering
    \includegraphics[width=\linewidth]{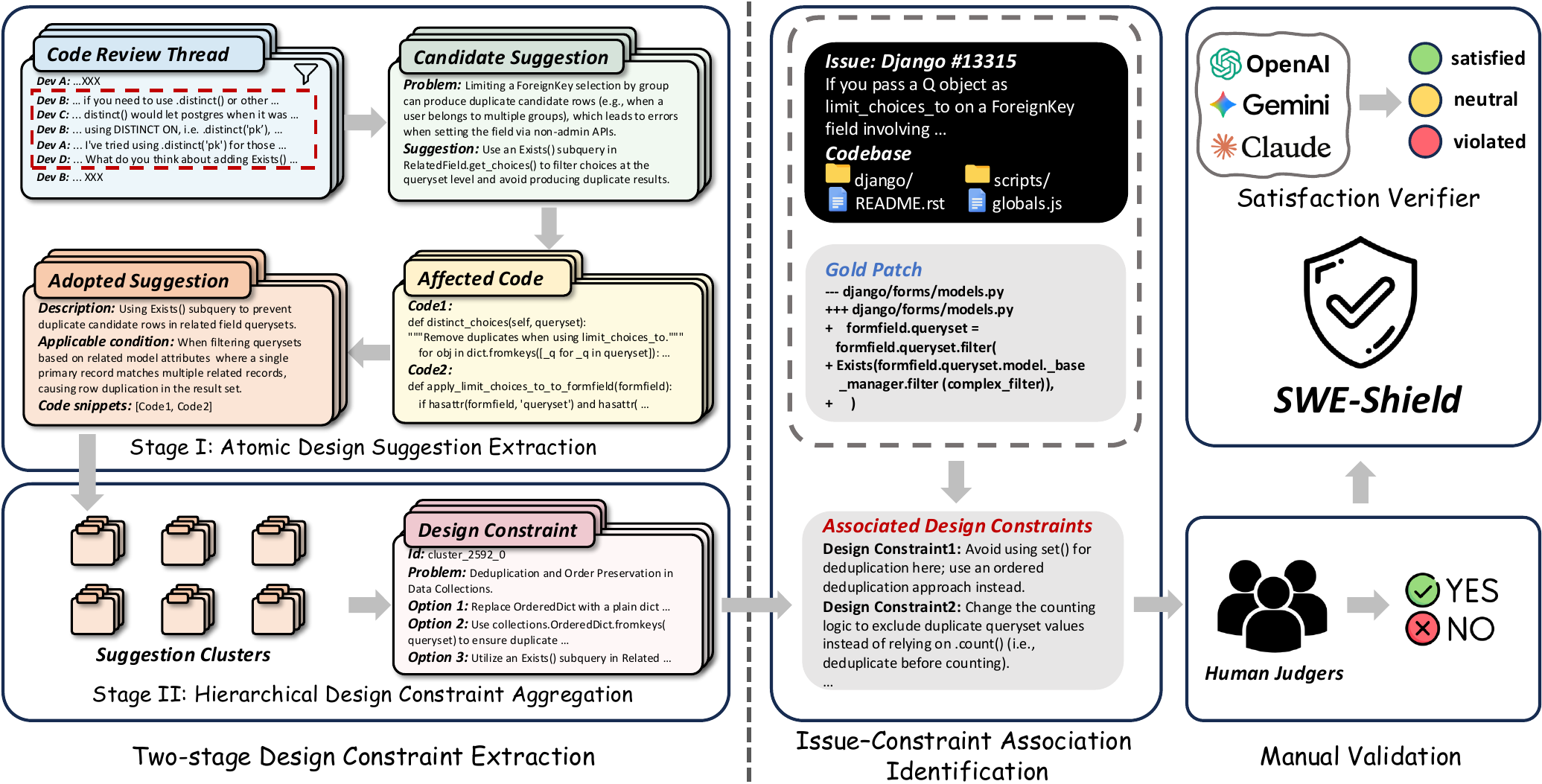}
    \caption{Overview of the construction pipeline of \bench}
    \label{fig:overview}
\end{figure*}

Figure~\ref{fig:overview} presents an overview of the construction pipeline of \bench. First, we extract design constraints from pull requests in each code repository using \tool, an LLM-based two-stage extraction approach. Next, we associate applicable design constraints with each target issue followed by manual validation. In addition, an accompanying patch verifier is provided based on LLMs-as-Judge to support the design satisfication verification for the generated patches.

A \textbf{design constraint} captures generalized design guidance while remaining grounded in scenario-specific design reasoning. It is represented as a structured object that links a design problem to a set of design options and the rationales that support them. In practice, SWE-SHIELD scopes constraints to project-level conventions, context-dependent design decisions, scenario-specific trade-offs, and cross-cutting concerns (e.g., error-handling protocols or API consistency) extracted from historical code review discussions, while excluding pure functional bug descriptions and simple style rules. Concretely, a design constraint consists of:

\begin{itemize}[leftmargin=15pt]
    \item A \textit{problem description}, which identifies the design issue or decision point being addressed.
    \item One or more \textit{design options}, each representing a possible way to address the problem. Each option is itself a structured representation that includes:
    \begin{itemize}[leftmargin=*]
        \item An \textit{option description}, which captures the suggested actions and the stated design rationale.
        \item An \textit{applicable condition}, which specifies when and in what contexts it is suitable to apply.
        \item Reference \textit{code snippets} associated with the option, which ground the option and its rationale in concrete code-level changes.
    \end{itemize}
\end{itemize}

An example of a design constraint is shown in the box labeled \emph{Design Constraint} in Figure~\ref{fig:overview}.

% Rationales are drawn directly from developer-authored pull request reviews and discussion threads and preserve their original context and source identifiers. They reflect how developers justify design constraints in specific scenarios, without assuming global validity or completeness.

% The construction of design constraints follows strict rules to maintain fidelity to the original rationale. First, every design option must be supported by one or more rationales; no option may be introduced without explicit textual justification. Second, source identifiers are preserved verbatim, ensuring end-to-end traceability from each design constraint back to the underlying review discussions. Third, the construction process is conservative: no new rationales, conditions, or applicability scopes are introduced beyond what can be directly inferred from the original discussions.

% This representation ensures that design constraints remain grounded in real developer reasoning while being structured in a form that supports systematic analysis, auditing, and application during issue resolution and design constraint evaluation.

\subsection{\tool: A Two-stage Design Constraint Extraction Approach}
\label{sec:extraction}
Given a set of pull requests from a code repository, \tool extracts design constraints through a two-stage process powered by LLMs. In the first stage, \tool analyzes individual code review threads in pull requests to \emph{decompose} them into \textit{atomic design suggestions}. These concrete design suggestions and rationales are distilled from noisy review artifacts such as code review comments and discussion threads. In the second stage, \tool analyzes the extracted design suggestions to \emph{recompose} them into design constraints. Specifically, suggestions that address the same design problem are grouped, contrasted, and aggregated into design options, together with their applicable conditions. Through this decomposition-and-recomposition process, \tool captures generalized design guidance grounded in recurring design reasoning across pull requests.

\subsubsection{Stage I: Atomic Design Suggestion Extraction}
In pull requests, code review threads often contain a large amount of noisy information, of which only a small portion concerns design suggestions and their associated rationales. Moreover, complete design suggestions are frequently distributed across multiple comments through multi-turn discussions and clarifications. In some cases, suggested design changes are ultimately rejected and not reflected in the final patch, leaving no corresponding code modifications in the commit history. These characteristics make it difficult to directly identify and validate design suggestions from raw code review threads. To address these challenges, \tool extracts atomic design suggestions from each code review thread using a sliding-window analysis over review comments, followed by validation against the commit history of the pull request to determine whether a suggested design was ultimately adopted.

\textbf{Sliding-Window Construction.} 
A naive approach to extracting design suggestions is to process all the normalized comments within a single LLM prompt. However, many code review threads involve long, multi-turn discussions that may span hundreds of comments. Processing such discussions in a single prompt often exceeds the effective context window of LLMs and leads to the well-known \emph{lost-in-the-middle} phenomenon~\cite{liu2024lost}, where salient information is diluted by surrounding noise. To mitigate this issue, \tool adopts a \textit{sliding-window strategy} that segments the comments in a thread into smaller windows. This approach is motivated by the observation that consecutive comments are often topically coherent, collectively discussing a specific design issue, concern, or implementation alternative within a localized span. By processing each window separately, the LLM can focus on the relevant discussion while still capturing local context and the logical relationships among comments.
Formally, given a code review thread consisting of an ordered list of comments $PR = [c_1, c_2, \ldots, c_n]$, \tool traverses the list using a window size $w$ and a step size $s = w$, , producing a sequence of non-overlapping comment windows. Each window is defined as $W_i = [c_{1+(i-1)s}, \ldots, c_{w+(i-1)s}]$. The window size $w$ reflects a trade-off between two competing factors. Smaller windows risk omitting dependencies across adjacent comments, while larger windows increase the likelihood of context dilution and reduced extraction precision. In practice, we find that $w = 6$ provides a reasonable balance, preserving local conversational coherence while remaining within the effective context limits of LLMs.

\textbf{LLM-based Suggestion Summarization.}
Each comment window $W_i$ is processed by an LLM to extract design suggestions contained in the comments. To enrich contextual understanding, the immediately preceding window $W_{i-1}$ is also provided as input when $i > 1$; however, the LLM uses it only as background context and does not directly incorporate its content into the final summarized suggestions. 
% The prompt template utilized for suggestion summarization is illustrated in Figure~\ref{fig:prompt_suggestion_extraction}.
The prompt template used for suggestion summarization is provided in our replication package.
This template is structured as a five-step chain of thought. Initially, the model is instructed to identify all problems explicitly stated within the dialogue. Subsequently, it excludes those problems that are purely procedural or process-oriented (\eg ``please rebase''). An extraction step is then conducted to derive the core suggestions and the corresponding reasons provided by the participants. Finally, the model is required to verify the source of its generated content, ensuring that its predictions are grounded solely in the dialogue text.

\textbf{Suggestion Adoption Verification.}
Not all suggested design actions are ultimately adopted in the final solution. Some suggestions are considered as alternatives, rejected during review, or remain unresolved. To determine whether an extracted suggestion is ultimately adopted, \tool verifies its semantically relevance to subsequent code modifications in the commit history of the corresponding pull request. Specifically, the verification proceeds as follows.

Given an extracted suggestion, \tool first locates the concrete source file and code lines that the suggestion refers to. This information can be recovered from the structure of code review threads in pull requests, which typically begin with a comment that directly quotes a range of code lines introduced in the initial patch. We denote this referenced code region as
$C_{\text{sugg}} = [l_1, l_2, \ldots, l_m]$.
\tool then performs lightweight, diff-based code correspondence tracing between the initial and final versions of the patch to determine whether the suggestion is reflected in subsequent code changes. Let $f_{\text{init}}$ and $f_{\text{final}}$ denote the source file versions corresponding to the initial patch and the final patch, respectively. \tool applies a standard diff tool (\eg difflib\footnote{https://docs.python.org/3/library/difflib.html}) to compute their differences:
$$
\textit{Diffs} = \textsc{compute-diff}(f_{\text{init}}, f_{\text{final}}).
$$
Each diff hunk in \textit{Diffs} includes metadata that specifies the affected line ranges in both versions. For example, a diff header of the form ``\texttt{@@ -144,6 +145,14 @@}'' indicates that 6 lines (144--149) in the initial version are deleted and replaced by 14 lines (145--158) in the final version. We denote the deleted and added code lines for a diff hunk as $C_{\text{del}}$ and $C_{\text{add}}$, respectively.
If $C_{\text{sugg}}$ overlaps with $C_{\text{del}}$, \tool treats the corresponding diff hunk as potentially relevant to the suggestion and adds it to a candidate set $R$. After collecting all candidate diffs in $R$, \tool constructs two aligned code snippets that reflect the code state before and after the suggestion. Specifically, it identifies the minimum and maximum line numbers across $C_{\text{sugg}}$ and all $C_{\text{del}}$ in $R$, and slices the corresponding range from $f_{\text{init}}$ as the \emph{before-suggestion} code snippet $C_{\text{before}}$. Similarly, it identifies the minimum and maximum line numbers across all $C_{\text{add}}$ in $R$ and slices the corresponding range from $f_{\text{final}}$ as the \emph{after-suggestion} code snippet $C_{\text{after}}$.
If no candidate diff is found, \ie $R = \varnothing$, the suggestion is deemed \emph{non-adopted}, as no corresponding code modification can be traced. 
% \chong{If having time, we can add an algorithm.}

% To tolerate line drift across commits—a common issue when files are modified between review and merge—we employ a multi-stage code location strategy. First, we attempt \textbf{fuzzy matching} to find appearance of the code lines in $C$ in the final source file content. The fuzzy matching algorithm computes a similarity score between the target snippet and candidate regions in the file, using a sliding window approach centered around the original line numbers from the review thread. We set a similarity threshold of 0.6 to balance precision and recall: lower thresholds risk false matches, while higher thresholds may miss valid but slightly modified code regions. When a match is found above the threshold, we extract the matched region plus surrounding context (typically 15 lines before and after) to provide sufficient code context for downstream analysis. If fuzzy matching fails, we fall back to a \textbf{diff-based range strategy}: we parse the diff hunk metadata to extract the current line range after all commits, and use that range (extended by 50 lines of context) as a secondary anchor point. This two-stage approach ensures that we can locate code even when significant refactoring has occurred between review and merge, while maintaining high precision by preferring semantic similarity over structural heuristics.

To determine whether the change from $C_{\text{before}}$ to $C_{\text{after}}$ genuinely implements the suggested design intent, \tool applies an LLM-based semantic checker. The checker evaluates whether $C_{\text{before}}$ violates the suggestion and whether $C_{\text{after}}$ satisfies it, based on the suggestion's stated reasoning and rationales. If the code change aligns with the intended design direction, \tool labels the suggestion as \emph{adopted}. During this determination, \tool also generates an \textit{applicable condition} for suggestions judged as \emph{adopted}, summarizing when and under what circumstances the suggested design choice should be applied. 
% The Adoption Verification prompt is presented in Figure~\ref{fig:prompt_adoption_verification}. 
The Adoption Verification prompt is provided in our replication package.
It incorporates a structured chain of thought consisting of four sequential steps. In the initial two steps, the model examines the provided core problem and the suggestion with corresponding code changes, with a focus on comprehending the relevant contextual background. Subsequently, it determines whether an adoption has taken place. Finally, the model identifies any supplementary conditions to ensure that the prerequisite for implementing the suggestion has not been overlooked.

% \begin{figure}[h] % h:当前位置, t:页顶, b:页底, p:独立一页
%     \centering % 居中对齐
%     \includegraphics[width=0.9\textwidth]{picture/prompt_adoption_verification.png} % 缩放至正文宽度的80%
%     \caption{Prompt Template for Adoption Verification} % 图片下方的文字说明
%     \label{fig:prompt_adoption_verification} % 用于文中引用的标签
% \end{figure}

For suggestions labeled as adopted, $C_{\text{before}}$ and $C_{\text{after}}$ are retained as reference code to ground the suggestion in concrete implementation changes. Importantly, non-adopted suggestions are not discarded at this stage. Instead, they are preserved as supporting references for subsequent synthesis, where they may contribute the synthesis of alternative design options, historical trade-offs, or applicable conditions within higher-level design constraints.

\subsubsection{Stage II: Hierarchical Design Constraint Aggregation}
\label{sec:hierarchically_synthesizing_design_decisions}

Given a collection of extracted design suggestions, \tool aggregates them to synthesize higher-level design constraints using LLMs. Directly presenting all suggestions to an LLM in a single prompt is impractical, as the number of suggestions often exceeds the effective context window and causes the model to lose focus, resulting in incoherent grouping or shallow abstractions. To address this limitation, \tool adopts a hierarchical aggregation strategy that incrementally groups related suggestions and synthesizes design constraints at appropriate levels of abstraction.

\textbf{Similarity-based Suggestion Clustering.}
\tool first organizes design suggestions into a hierarchical clustering structure based on a unified similarity measure that combines semantic similarity and structural dependency. Semantic similarity captures whether two suggestions express related design concerns, while structural dependency reflects their historical proximity in the development process, such as whether they originate from the same comment thread or pull request. Semantic similarity serves as the primary signal, with structural dependency providing contextual refinement when semantic cues alone are ambiguous.
Specifically, for semantic similarity, \tool embeds the problem descriptions and suggestion texts of design suggestions using a sentence transformer model (\texttt{all-MiniLM-L6-v2})~\cite{DBLP:conf/emnlp/ReimersG19}, producing dense vector representations that capture semantic meaning. The problem description is weighted more heavily (0.8) than the suggestion text (0.2), reflecting that problem formulations are more stable indicators of design intent than individual solution proposals. \tool computes pairwise cosine similarities between embeddings to obtain a semantic similarity matrix. For structural dependency, \tool computes similarity scores based on proximity in the review process. Suggestions from the same review thread receive the highest similarity score (1.0), followed by suggestions from the same review (0.7) and the same pull request (0.3). \tool further applies small bonuses (+0.2) when suggestions reference the same file path or occur within a short time window, capturing spatial and temporal locality. \tool computes the final combined similarity as
$
s_{\text{combined}} = w_s \cdot s_{\text{semantic}} + w_t \cdot s_{\text{structural}},
$
where $w_s = 0.8$ and $w_t = 0.2$. This weighting ensures that semantic alignment dominates clustering decisions. \tool employs a hierarchical clustering algorithm~\cite{DBLP:journals/jacm/Cohen-AddadKMM19} to organize the extracted design suggestions into candidate groups. Specifically, \tool first computes pairwise similarities among all suggestions and then iteratively merges the two most similar groups until a hierarchical dendrogram is formed. Each leaf corresponds to an individual suggestion, and each internal node represents the merge of two suggestion groups. To obtain candidate groups at a controlled granularity, we cut the tree with a similarity threshold $\tau$ ($0.6$ in our implementation): internal nodes with similarity $\geq \tau$ are retained as candidate groups, while merges under $\tau$ are prevented. This yields suggestion groups of varying sizes without requiring a predefined number of clusters.

\textbf{LLM-based Constraint Synthesis.}
\tool synthesizes design constraints by performing a post-order traversal of the clustering tree. 
Specifically, \tool first transforms each leaf node of the extracted suggestion tree into a design constraint with a single option. Then, for each internal node, \tool supplies the child design constraints to an LLM and instructs it to abstract their shared design intent while strictly preserving the original meaning and scope. The LLM is guided to determine whether the child constraints should be (i) merged into a single design constraint that captures a common underlying concern at a higher level of abstraction, or (ii) split into multiple independent design constraints if they address distinct design problems.
When constraints are merged, \tool refines their options by eliminating redundancies, consolidating semantically similar descriptions and compatible conditions, and aggregating all source identifiers. This reduces redundancy while preserving distinct design intents and traceability.
After processing an internal node, it is replaced either by a new leaf node if all child constraints are merged into a single constraint, or by a set of leaf nodes representing newly formed constraints reorganized from the original children. This ensures that, during post-order traversal, each internal node operates on the most up-to-date child constraints.
Throughout this process, \tool enforces a strict rule to mitigate hallucinations: all synthesized design options must be grounded in one or more original suggestions and maintain explicit traceability to their reference code and review comments.
Each design constraint is represented by \tool as a structured abstraction that includes a normalized problem description, a set of alternative design options, explicit traceability links to the original suggestions and code snippets, and minimal metadata required to preserve hierarchical relationships.

\subsection{Issue–Constraint Association Identification}
Given an issue in a code repository resolved by a merged pull request, we attempt to associate it to relevant design constraints mined from the repository's historical code review threads, serving as \emph{implicit design considerations} beyond functional correctness measured by test cases. In this way, each issue linked to constraints is treated as a \emph{\textbf{design-aware issue resolution task}}.

Because design reasoning in real-world repositories is often mixed within comments and scattered across review threads and pull requests, we employ two complementary association channels. Both aim to link design constraints to a target issue but rely on different forms of evidence. Channel~A exploits explicit traceability between the issue and the code review threads of its resolving pull request, while Channel~B performs semantic matching to retrieve potentially relevant design constraints from code review discussions in other pull requests across the repository's history.

\begin{itemize}[leftmargin=15pt]
    \item \textit{Channel A: Association via Explicit Traceability.}
    We leverage explicit traceability information preserved during the design constraint extraction process. Each issue and pull request is associated with a set of extracted design suggestions, and each design constraint maintains provenance links to the suggestions that support its options. To establish the association, we collect identifiers of design suggestions linked to the target issue and its resolving pull request, and build an inverted index that maps suggestion identifiers to the design constraints that reference them. Using this index, we retrieve all design constraints that cite any of the collected suggestions from the resolving pull request of the target issue. This channel provides \emph{high-precision} associations, as each retrieved design constraint is grounded in at least one human-authored design discussion directly tied to the issue’s surrounding code review context.

    \item \textit{Channel B: Association via Semantic Matching.}
    When explicit traceability is unavailable, we supplement Channel~A by linking the issue to broader, repository-wide design constraints. These constraints may be supported by discussions spread across other pull requests that do not directly resolve the target issue. To establish this association, we measure the semantic similarity between the resolving patch and each design constraint in $\mathcal{D}$. Each design constraint is represented by its normalized problem and option descriptions, while the patch is represented as a set of natural-language \emph{change intents}, generated by applying an LLM to analyze the code diffs and extract explicit design and implementation decisions along multiple dimensions (\eg performance, reliability, and maintainability). Both design constraints and change intents are embedded into a shared vector space using the sentence transformer model. The relevance of a design constraint to the issue is determined by the maximum cosine similarity between the constraint's representation and any of the patch's change-intent embeddings. This allows us to identify constraints that are most semantically aligned with the changes introduced by the patch, even when there is no direct traceability.

\end{itemize}

% \paragraph{Dataset Artifact.}
% Each benchmark instance is represented as
% \[
% (\mathcal{R}_{\text{base}}, I, P^\star, T, \mathcal{D}_I),
% \]
% where $\mathcal{R}_{\text{base}}$ denotes the base repository state, $I$ is the issue description, $P^\star$ is the developer-accepted patch, $T$ is the test suite, and $\mathcal{D}_I$ is the set of associated design constraints. Each design constraint in $\mathcal{D}_I$ includes its alternative design options, supporting rationales with provenance, and applicability conditions.

% This formulation enables evaluation along two orthogonal dimensions:
% \begin{itemize}
%     \item \emph{functional correctness}, measured by passing the test suite $T$; and
%     \item \emph{design alignment}, measured by consistency with the associated design constraints $\mathcal{D}_I$.
% \end{itemize}

\subsection{An Accompanying Patch Verifier: LLMs-as-Judge for Design Satisfaction} 
Unlike functional correctness, which can be directly verified by executing test cases, determining whether a generated patch satisfies design constraints requires semantic reasoning. Motivated by the recent success of LLMs-as-judge in deterministic evaluation tasks~\cite{Szymanski2024}~\cite{Ye2024} ~\cite{Chen2024}, we employ an LLM-based judge with a voting mechanism to assess whether a patch aligns with a given design constraint. The verification procedure is formulized as:
% $\{ \text{Satisfied}, \text{Neutral}, \text{Violated} \} \leftarrow \textsc{verify}(patch, constraint).$
\begin{equation}
\small\{Satisfied, Neutral, Violated\} \leftarrow \textsc{verify}(patch, constraint).
\end{equation}
Specifically, given a patch and a design constraint, the LLM compares the patch against the constraint's problem description, alternative options, and associated applicable conditions and rationales. It evaluates whether the changes introduced by the patch fulfill the intended design intent and context behind the constraint, taking into account both structural and behavioral aspects of the modification.
Given a patch and a design constraint, we adopt an LLM-as-judge protocol to assess whether the patch respects the design constraint and applicability context. The evaluation prompt consists of three components: the issue context, the set of design options and the agent-generated patch. The LLM performs a two-step analysis by first determining whether the patch matches the applicability condition of each option, and then classifying the option as \textit{Satisfied}, \textit{Violated}, or \textit{Neutral}. A constraint is considered \textit{Satisfied} if the patch adopts the prescribed design option, \textit{Violated} if it contradicts the option’s requirements, and \textit{Neutral} if the option is not applicable to the concrete patch changes. \textit{Neutral} is needed because many constraints are \emph{conditionally applicable}, meaning their relevance depends on how the patch modifies the code (e.g., whether a specific API is used). During manual validation, we preserve these relevant but conditionally applicable constraints to enable a more comprehensive evaluation. Accordingly, we include a \textit{Neutral} outcome for constraints that are considered but do not apply to the concrete patch changes. Each evaluation returns a structured JSON output with reasoning and a confidence score. To improve robustness, we employ three independent LLMs to evaluate each patch–constraint pair in parallel and determine the final label via majority voting, where \textit{Satisfied} or \textit{Violated} requires agreement from at least two models, and \textit{Neutral} is assigned otherwise.

\section{Implementation}
In this section, we present the details of the benchmark construction, along with the manual reliability validation of several key components.

\subsection{Construction Pipeline}

We apply the construction pipeline to two existing issue resolution benchmarks to derive a new benchmark \bench with design constraints. 

\textbf{Existing Benchmark Selection.} We select two representative and widely adopted issue resolution benchmarks as the foundation: SWE-bench~\cite{DBLP:conf/iclr/JimenezYWYPPN24} (using its verified subset) and SWE-bench Pro~\cite{DBLP:journals/corr/abs-2509-16941}. Based on these two benchmarks, we construct two corresponding variants of our benchmark, \benchVerified and \benchPro, respectively.

\textbf{Repository and Issue Selection.}
We further filter repositories and issues from the selected benchmarks using the following steps. First, we rank all repositories by the number of associated issues in descending order. Second, we collect all issues from repositories that contain more than 40 issues, ensuring sufficient issue diversity and representativeness.
Following this process, we obtain a total of 618 issues from the two benchmarks, including 306 from two repositories in SWE-bench-Verified and 312 from four repositories in SWE-bench-Pro.

\textbf{Design Constraint Extraction.}
To extract comprehensive design constraints for each issue, we augment the context of each target issue with additional related issues from the same repository. Specifically, for each issue, we retrieve its corresponding pull request (PR) and identify the top-20 most relevant PRs based on a combination of PR title similarity and patch-level file path similarity. The issues associated with these PRs are then jointly used with the target issue as input to the design constraint extraction process described in Section~\ref{sec:extraction}. Using this procedure, we initially extract 10,885 design constraints, including 4,695 from SWE-bench-Verified and 6,190 from SWE-bench-Pro.

\textbf{Issue-Constraint Association.}
After constructing the associations, 2,458 design constraints are associated to 648 issues, including 937 constraints for 306 issues in SWE-bench-Verified and 1,521 constraints for 342 issues in SWE-bench-Pro.

\subsection{Manual Validation}
We perform three manual validation processes targeting the key components: the \tool extractor, issue–constraint association identification, and the LLM-as-Judge patch verifier.

\textbf{Reliability of \tool.} 
To further corroborate the practical validity of \tool, we conduct a manual evaluation on the extracted design constraints involving two domain experts—each with over five years of professional development experience. We first randomly sample 374 constraints from the 10,885 extracted constraints from SWE-bench-Verified and SWE-bench-Pro, using a common statistical sampling method~\cite{ahmad2017determining}. The independent annotations demonstrated substantial inter-rater reliability (Cohen's kappa of 0.74), with 90.4\% of the sampled constraints ultimately verified as valid for code patching.

\textbf{Reliability of Association Identification.}
This step directly yields the issue resolution tasks in the benchmark. We recruit two annotators, each with more than four years of experience in Python, Java, and C/C++ development, to label all 2,458 identified issue–constraint associations. Each instance is independently reviewed by both annotators, and any disagreements are resolved by a third annotator to ensure consistency and reliability.

Annotators evaluate each instance according to the following objective criteria. A design constraint is considered \emph{associated} with an issue instance only if all of the following conditions are satisfied:
(a) \emph{Constraint quality:} the constraint options are supported by explicit evidence and are sufficiently specific to be verified against code changes;
(b) \emph{Issue relevance:} the constraint addresses a design concern relevant to the issue, \ie its condition is likely to hold for the issue instance, or the affected code pattern described by the constraint matches the entities modified by the reference patch (\eg similar implementation patterns).

The two annotators achieve a Cohen's kappa of 0.7783, indicating substantial agreement. As a result, 1,787 issue–constraint associations are labeled as high quality and retained in the final benchmark. We exclude all issues for which no valid design constraints are extracted. Consequently, \bench comprises 495 issues associated with 1,787 high-quality design constraints.

\textbf{Reliability of Patch Verifier.}
In the LLM-as-judge verifier, we adopt a majority-voting scheme over three state-of-the-art LLMs and assess their internal agreement. Specifically, we compute the proportion of cases in which at least two of the three models agree, which averages 95.25\% across all instances. This high level of agreement indicates strong consistency among LLM-based judges.

To further validate the LLM-based verifier against human judgment, we randomly sample 318 patches from 1,842 patches generated by the evaluated agents (see Section\ref{evaluation}) on \bench. Two human experts, each with more than five years of development experience, independently assess the judgments produced by the LLM ensemble~\cite{ahmad2017determining}. The comparison yields a consistency rate of 80.8\% between human annotations and the verifier judgments, with a Cohen's kappa of 0.7934, indicating substantial agreement. These results suggest that LLMs can serve as reliable proxies for human evaluation in assessing design constraint compliance.

\subsection{Benchmark Characteristics.}
Table~\ref{table:dataset} summarizes the statistics of the \bench benchmark. Overall, \bench comprises 495 issue resolution tasks associated with 1,787 high-quality design constraints. 

To further characterize the benchmark, we analyze patch size, and language diversity.For patch size, SWE-Bench-Verified remains small-scale, averaging 12.99 changed lines (max 156), with 303/306 issues within 0–99 lines. In contrast, SWE-Bench-Pro involves much larger modifications, averaging 197.92 lines (max 2,028), with many issues exceeding 100 lines, reflecting significantly higher modification complexity.
Finally, SWE-Bench-Verified is limited to Python , whereas SWE-Bench-Pro spans multiple languages, demonstrating greater diversity and broader applicability.

% \todo{we can present more statistical information, such as, the distribution of constraint numbers for each issue, languages, repo LoC, patch size, etc.}

% \begin{table}[t]
%   \centering
%   \caption{Distribution of repositories and issues in SWE-Shield}
%   \label{table:dataset}
%   \small
%   \setlength{\tabcolsep}{8pt}
%   \begin{tabularx}{\columnwidth}{l l c c}
%     \toprule
%     \textbf{Benchmark} & \textbf{Repository} & \textbf{\#Issues} &
%     \makecell{\textbf{\#Assoc.}\\\textbf{Constraints}} \\
%     \midrule
%     \multirow{2}{*}{\benchVerified{}} 
%       & django & 182 & 590 \\
%       & sympy   & 54  & 132 \\
%     \midrule
%     \multirow{4}{*}{\benchPro{}} 
%       & ansible             & 84 & 480 \\
%       & teleport      & 73 & 331 \\
%       & flipt              & 62 & 149 \\
%       & openlibrary & 40 & 105 \\
%     \bottomrule
%   \end{tabularx}
% \end{table}

\begin{table}[t]
  \centering
  \caption{Distribution and Characteristics of Repositories and Issues in SWE-Shield}
  \label{table:dataset}
  \footnotesize % 保持较小字号确保在双栏中留有余量
  \setlength{\tabcolsep}{4pt} % 适当放宽列间距，因为少了一列
  \begin{tabularx}{\columnwidth}{@{} l X c c c l @{}}
    \toprule
    \textbf{Benchmark} & \textbf{Repository} & \textbf{\#Iss.} & \textbf{\#Cons.} & \makecell{\textbf{Avg.}\\\textbf{Patch Lines}} & \textbf{Lang.} \\
    \midrule
    \multirow{2}{*}{\benchVerified{}} 
      & django & 182 & 590 & \multirow{2}{*}{\makecell{12.99\\(max 156)}} & \multirow{2}{*}{Python} \\
      & sympy  & 54  & 132 &  & \\
    \midrule
    \multirow{4}{*}{\benchPro{}} 
      & ansible     & 84 & 480 & \multirow{4}{*}{\makecell{197.92\\(max 2,028)}} & \multirow{4}{*}{\makecell{Multi\\(Py,Go,etc)}} \\
      & teleport    & 73 & 331 &  & \\
      & flipt       & 62 & 149 &  & \\
      & openlibrary & 40 & 105 &  & \\
    \bottomrule
  \end{tabularx}
\end{table}

\section{Empirical Study} 
\label{evaluation}
Based on \bench{}, we conduct the first study that evaluates existing LLM-based agents on their awareness of, and compliance with, design constraints during issue resolution. Specifically, we answer the following research questions:

% 最终采样了 60 个选项用于评估

\begin{itemize}[leftmargin=10pt, topsep=5pt]
\item \textbf{RQ1 (Effectiveness in Design-Aware Resolution)}: To what extent do existing LLM-based agents comply with design constraints during issue resolution?

\item \textbf{RQ2 (Correlation between Correctness and Satisfaction)}: What is the relationship between functional correctness and design satisfaction?

\item \textbf{RQ3 (Comparison across Foundation Models):} How do different LLMs compare in terms of design satisfaction?

\item \textbf{RQ4 (Investigation on Design Satisfaction Improvement)}: Can providing relevant design-constraint guidance improve design satisfaction?
\end{itemize}

\subsection{Experimental Setup}

% \subsubsection{Benchmark Info.}
% \zq{About evaluation used dataset, e.g., constructed, \textbf{size}, ..}

\subsubsection{Studied LLM-based Agents}
We study the state-of-the-art LLM-based agents, including SWE-agent~\citep{DBLP:conf/nips/YangJWLYNP24}, Live-SWE-agent~\citep{DBLP:journals/corr/abs-2511-13646}, Lingxi-v1.5~\citep{DBLP:journals/corr/abs-2510-11838}, and the Sonar Foundation Agent~\citep{ruan2025sonarfoundationagent}. These agents have achieved high effectiveness on recent issue resolution leaderboards (\ie SWE-bench-verified~\citep{DBLP:conf/iclr/JimenezYWYPPN24} and SWE-Bench Pro~\citep{DBLP:journals/corr/abs-2509-16941}). The selected agents are powered by competitive frontier LLMs, including Kimi-K2, GPT-5, Claude-Sonnet-4.5, Gemini-2.5-Pro, and Gemini-3.0-Pro.

\begin{table*}[t]
\centering
\caption{Overall performance and outcome distribution of different agents on \bench{}}
\label{tab:rq1_overall_performance}
\resizebox{\linewidth}{!}{
\begin{tabular}{llcccc>{\columncolor{gray!20}}c ccc}
\toprule
\textbf{Dataset} & \textbf{Agent} 
& \textbf{DSR (\%)} & \textbf{DVR (\%)} & \textbf{DNR (\%)} & \textbf{Pass (\%)} 
& \cellcolor{white}\textbf{P\&S (\%)}  & \textbf{P\&V}(\%) & \textbf{F\&S}(\%) & \textbf{F\&V}(\%) \\
\midrule
\multirow{4}{*}{\benchPro}
 & SWE-agent (Gemini-2.5-Pro)
 & {41.50} & 39.92 & 18.58 & \underline{13.44}
 & \underline{4.74} & \underline{8.70} & \textbf{36.76} & \textbf{49.80} \\
& SWE-agent (Kimi-K2)
 & {40.71} & \textbf{43.08} & 16.21 & 18.18
 & 7.51 & 10.67 & 33.20 & 48.62 \\
 & SWE-agent (GPT-5)
 & \underline{39.13} & 41.11 & \textbf{19.76} & 30.43
 & 13.83 & 16.60 & \underline{27.27} & 42.29 \\
 & SWE-agent (Claude-Sonnet-4.5)
 & \textbf{50.20} & \underline{37.15} & \underline{12.65} & \textbf{42.69}
 & \textbf{22.53} & \textbf{20.16} & 27.67 & \underline{29.64} \\
\midrule
\multirow{3}{*}{\benchVerified}
 & Lingxi-v1.5 (Kimi-K2)
 & \underline{32.64} & \textbf{43.39} & 23.97 & \underline{70.25}
 & \underline{25.62} & \textbf{44.63} & \underline{7.02} & \textbf{22.73 }\\
 & Sonar Foundation Agent (Claude-Sonnet-4.5)
 & {39.17} & 36.67 & \textbf{24.17} & 73.75
 & 30.00 & 43.75 &\textbf{ 9.17} & 17.08 \\
 & Live-SWE-agent (Gemini-3.0-pro)
 & \textbf{42.80} & \underline{36.21} & \underline{20.99} & \textbf{76.95}
 & \textbf{34.57} & \underline{42.39} & 8.23 & \underline{14.81} \\
\bottomrule
\end{tabular}
}
\footnotesize\raggedright{
\emph{Note:} P/F denotes whether a patch passes/fails benchmark tests;
S/V denotes whether it satisfies/violates applicable design constraints.
Underlined values denote the minimum, whereas bold values denote the maximum.
}
\end{table*}

\subsubsection{Evaluation Metrics}
We evaluate issue resolution on \bench{} from two complementary perspectives:
\emph{functional correctness} and \emph{design satisfaction}.
Functional correctness is measured by \emph{Pass Rate}, while design satisfaction is characterized by three design-aware metrics that form a mutually exclusive partition over instances: \emph{Design Satisfaction Rate (DSR)}, \emph{Design Violation Rate (DVR)}, and \emph{Design Neutral Rate (DNR)}.

\textbf{Pass Rate.}
Following prior work~\citep{DBLP:conf/iclr/JimenezYWYPPN24, DBLP:journals/corr/abs-2509-16941}, an instance is considered \emph{passed} if the generated patch passes all predefined tests. \emph{Pass Rate} is the fraction of issues that pass the test cases.

\textbf{Design Satisfaction Rate (DSR).} To evaluate design satisfaction, we associate each issue instance $I_i$ ($i\in\{1,\ldots,N\}$) with a set of design constraints
$\mathcal{DC}_i=\{dc_{i,1},\ldots,dc_{i,m_i}\}$, where each $dc_{i,j}$ captures a project-specific design rule extracted from developer discussions, Given a generated patch $\hat{p}_i$, we judge each constraint along two dimensions:
(i) \emph{applicability} $\mathrm{app}(\hat{p}_i, dc_{i,j})\in\{0,1\}$, indicating whether the constraint is relevant to the patch context; and
(ii) \emph{satisfaction} $\mathrm{sat}(\hat{p}_i, dc_{i,j})\in\{0,1\}$, indicating whether the patch complies with the constraint.
Satisfaction is assessed only when $\mathrm{app}(\hat{p}_i, dc_{i,j})=1$.

DSR measures the fraction of instances whose patches satisfy \emph{all applicable} design constraints based on the following Equation:
\begin{equation}
\mathrm{DSR}=\frac{1}{N}\sum_{i=1}^{N}\mathbb{I}\!\left[\mathrm{Satisfied}(i)\right]
\end{equation}

Where $i$ indexes an issue instance, $N$ is the number of evaluated issues, and $\hat{p}_i$ is the
patch generated for issue $i$. $\mathcal{D}_i$ denotes the set of design constraints retrieved and
validated for issue $i$, and $dc$ denotes one design constraint in $\mathcal{D}_i$.
$\textit{app}(\hat{p}_i, dc)\in\{0,1\}$ indicates whether $dc$ is applicable to $\hat{p}_i$, and $\textit{sat}(\hat{p}_i, dc)\in\{0,1\}$ indicates whether $\hat{p}_i$ satisfies $dc$
when applicable. $\mathbb{I}[\cdot]$ is the indicator function.
%Formally, an instance is labeled \emph{satisfied} ifthere exists at least one applicable constraint and all applicable constraints are satisfied:

\begin{equation}
\begin{aligned}
A_i &\triangleq \{\, j \mid \mathrm{app}(\hat{p}_i, dc_{i,j})=1 \,\},\\
\mathrm{Satisfied}(i) &\triangleq
\left(A_i \neq \emptyset\right)\ \wedge\
\left(\bigwedge_{j\in A_i}\ \mathrm{sat}(\hat{p}_i, dc_{i,j})=1\right).
\end{aligned}
\end{equation}

%We then compute:

\textbf{Design Violation Rate (DVR).}
DVR captures instances where the patch violates at least one applicable design constraint.
% :
% \[
% \mathrm{Violated}(i) \triangleq
% \exists j:\ \mathrm{app}(\hat{p}_i, dc_{i,j})=1\ \wedge\ \mathrm{sat}(\hat{p}_i, dc_{i,j})=0,
% \qquad
% \mathrm{DVR}=\frac{1}{N}\sum_{i=1}^{N}\mathbb{I}\!\left[\mathrm{Violated}(i)\right].
% \]

\textbf{Design Neutral Rate (DNR).}
DNR captures instances for which none of the associated design constraints are applicable to the generated patch.

\subsection{RQ1: Effectiveness in Design-Aware Resolution}

% To assess whether agent-generated patches respect project-specific design constraints \emph{beyond} test-based functional correctness, 

Table~\ref{tab:rq1_overall_performance} demonstrates the performance of all studied LLM-based agents on both \benchPro and \benchVerified.
Overall, our studied agents demonstrate SOTA performance from the perspective of the Pass Rate. On \benchPro, the Pass Rate ranges from 13.44\% to 42.69\%. On \benchVerified, the performance is more remarkable, with the Pass Rate ranging from 70.25\% to 75.95\%, which means that these agents can fix nearly three-quarters of the issues. 

However, from the perspective of design satisfaction, the DSR data remains consistently low across both datasets, ranging from 39.13\% to 50.20\% on \benchPro and 32.64\% to 42.80\% on \benchVerified.
% Conversely, DVR remains high (37.15\%-43.08\% on \benchPro and 36.21\%-43.39\% on \benchVerified), indicating that design-constraint violations occur frequently among generated patches.
An illustrative example is SWE-agent with Claude-Sonnet-4.5; even with the best-performing DSR on \benchPro, there are still half of the generated patches that fail to adhere to the design constraints extracted from the original code repository, highlighting the lack of design awareness in existing issue-resolution agents.

We further orthogonalize the generated patches along two dimensions, whether the patches satisfy design constraints (S/V) and whether the benchmark tests are passed (P/F), and obtain four categories, P\&S, P\&V, F\&S, and F\&V, which represent the four possible combinations of design compliance and test results. Experimental results show a substantial drop from the original Pass Rate to P\&S, highlighting that generating a perfect patch that meets both design constraints and benchmark tests remains a significant challenge.

\summary{RQ1 shows a persistent gap between functional correctness and design satisfaction.
Across both datasets, although the studied agents achieve SOTA performance on Pass Rate, fewer than half of the generated patches are fully design-aligned (DSR $\leq$ 50.20\%), which highlights design satisfaction as an orthogonal, largely unsolved dimension not captured by test-based evaluation alone.}

\subsection{RQ2: Correlation between Correctness and Satisfaction}

% RQ2 examines whether test-based functional correctness serves as a reliable proxy for design satisfaction.
% Table~\ref{tab:rq2} reports the joint distribution of test outcomes (Pass/Fail) and design judgments (Satisfied/Violated) for instances where at least one design constraint is applicable; non-applicable cases are captured by DNR in Table~\ref{tab:rq1_overall_performance}. 
Table ~\ref{tab:rq2} further reports the statistical relationships between functional correctness (P/F on benchmark tests) and design satisfaction (S/V on design constraints). Constraints with a neutral status are excluded, as they are not applicable to the generated patch. The $p$-value~\cite{fisher1970statistical} is computed using a $\chi^2$ test of independence~\cite{pearson1900x}.
Cramér’s \cite{cramer1999mathematical} $V$ measures the effect size between functional correctness and design satisfaction.

% \textit{Outcome Breakdown.}
% Table~\ref{tab:rq2} shows the joint outcomes of test-based correctness (Pass/Fail) and design judgment (Satisfied/Violated) for each agent, with non-applicable cases accounted for separately by DNR (Table~\ref{tab:rq1_overall_performance}).
% Across both \benchPro{} and \benchVerified{}, test passing does not guarantee design compliance:
% a substantial portion of test-passing patches still violates at least one applicable design constraint (P\&V reaches 20.16\% on \benchPro{} and 42.39\%--44.63\% on \benchVerified{}).
% Moreover, P\&V is often comparable to, or larger than, F\&S, indicating that passing tests does not reliably reduce the risk of design violations.

% \textit{Joint outcome patterns.}
% Across both settings, a substantial fraction of test-passing patches still violates at least one applicable design constraint.
% On \benchVerified{}, P\&V remains consistently high (42.39\%--44.63\%) across agents, often exceeding P\&S (25.62\%--34.57\%), indicating that passing tests does not reliably reduce the risk of design violations.
% A similar pattern appears on \benchPro{}, where P\&V ranges from 8.70\% to 20.16\% and is comparable to P\&S for several agents, suggesting that functional success and design compliance frequently diverge.

\textit{Statistical Association.}
We further test whether test outcomes and design judgments are statistically associated using a $\chi^2$ test of independence, and report Cramér’s $V$ as an effect-size measure.
Across all agents and datasets, the association remains negligible (all $V \leq 0.1157$), and the $\chi^2$ tests are not significant in most settings.
These results suggest that test-based correctness provides little information about whether a patch complies with project-specific design constraints.

\textit{Mismatch Characteristics.}
We observe systematic mismatches in which patches pass the benchmark tests yet violate grounded design constraints.
Such violations often concern high-impact constraints, including security-relevant checks, error-handling protocols, and maintainability-related design boundaries, which unit tests do not explicitly exercise.
As a result, test-based evaluation may label these patches as fully successful while they introduce latent design erosion.

\textit{Implications.}
Overall, Table~\ref{tab:rq2} indicates that functional correctness and design satisfaction capture complementary, largely orthogonal dimensions of patch quality.
Therefore, optimizing and evaluating agentic issue resolution solely via test outcomes is insufficient; design-aware evaluation (\eg{} DSR/DVR/DNR) is necessary to reflect repository-specific requirements beyond the test suite.

\summary{\textit{Test passing is a poor proxy for design compliance.}
Test outcomes and design judgments exhibit negligible association (Cramér’s $V \leq 0.11$), and many patches pass tests while violating applicable design constraints, motivating explicit design-aware evaluation beyond functional correctness.}

% \begin{table}[t]
% \centering
% \caption{Statistical relationship between functional correctness and design satisfaction on \bench{}}
% \label{tab:rq2}
% \small
% \resizebox{0.7\linewidth}{!}{
% \begin{tabular}{lcc}
% \toprule
% \textbf{Agent} & \textbf{$p$-value} & \textbf{Cram\'er's V} \\
% \midrule
% SWE-agent (Kimi-K2)                 & 1.0000 & 0.0000 \\
% SWE-agent (Gemini-2.5-Pro)          & 0.5468 & 0.0379 \\
% SWE-agent (GPT-5)                   & 0.4290 & 0.0497 \\
% SWE-agent (Claude-Sonnet-4.5)       & 0.5611 & 0.0365 \\
% \midrule
% Live-SWE-agent (Gemini-3.0-pro)     & 0.2858 & 0.0685 \\
% Lingxi-v1.5 (Kimi-K2)               & 0.0718 & 0.1157 \\
% Sonar Foundation Agent (Claude-Sonnet-4.5) & 0.5133 & 0.0422 \\
% \bottomrule
% \end{tabular}
% }

% \end{table}

\begin{table}[tbp]
\centering
\caption{Statistical relationship between functional correctness and design satisfaction on SWE-Shield.}
\label{tab:rq2}
\small
\setlength{\tabcolsep}{6.5pt}
\renewcommand{\arraystretch}{1.08}
\begin{tabular}{lrr}
\toprule
\textbf{Agent} & \textbf{$p$-value} & \textbf{Cram\'er’s $V$} \\
\midrule
SWE-agent (Kimi-K2) & 1.0000 & 0.0000 \\
SWE-agent (Gemini-2.5-Pro) & 0.5468 & 0.0379 \\
SWE-agent (GPT-5) & 0.4290 & 0.0497 \\
SWE-agent (Claude-Sonnet-4.5) & 0.5611 & 0.0365 \\
\midrule
Live-SWE-agent (Gemini-3.0-pro) & 0.2858 & 0.0685 \\
Lingxi-v1.5 (Kimi-K2) & 0.0718 & 0.1157 \\
Sonar Foundation Agent (Claude-Sonnet-4.5) & 0.5133 & 0.0422 \\
\bottomrule
\end{tabular}
\end{table}

\vspace{-3mm}
\subsection{RQ3: Comparison across Foundation Models}

RQ3 examines how different foundation models vary in recognizing and complying with project-specific design constraints.
We focus on \benchPro{}, where all settings adopt the same agent framework (\textsc{swe-agent}), enabling a model-centric comparison.

\begin{figure}[bp]
  \centering
  \includegraphics[width=\linewidth]{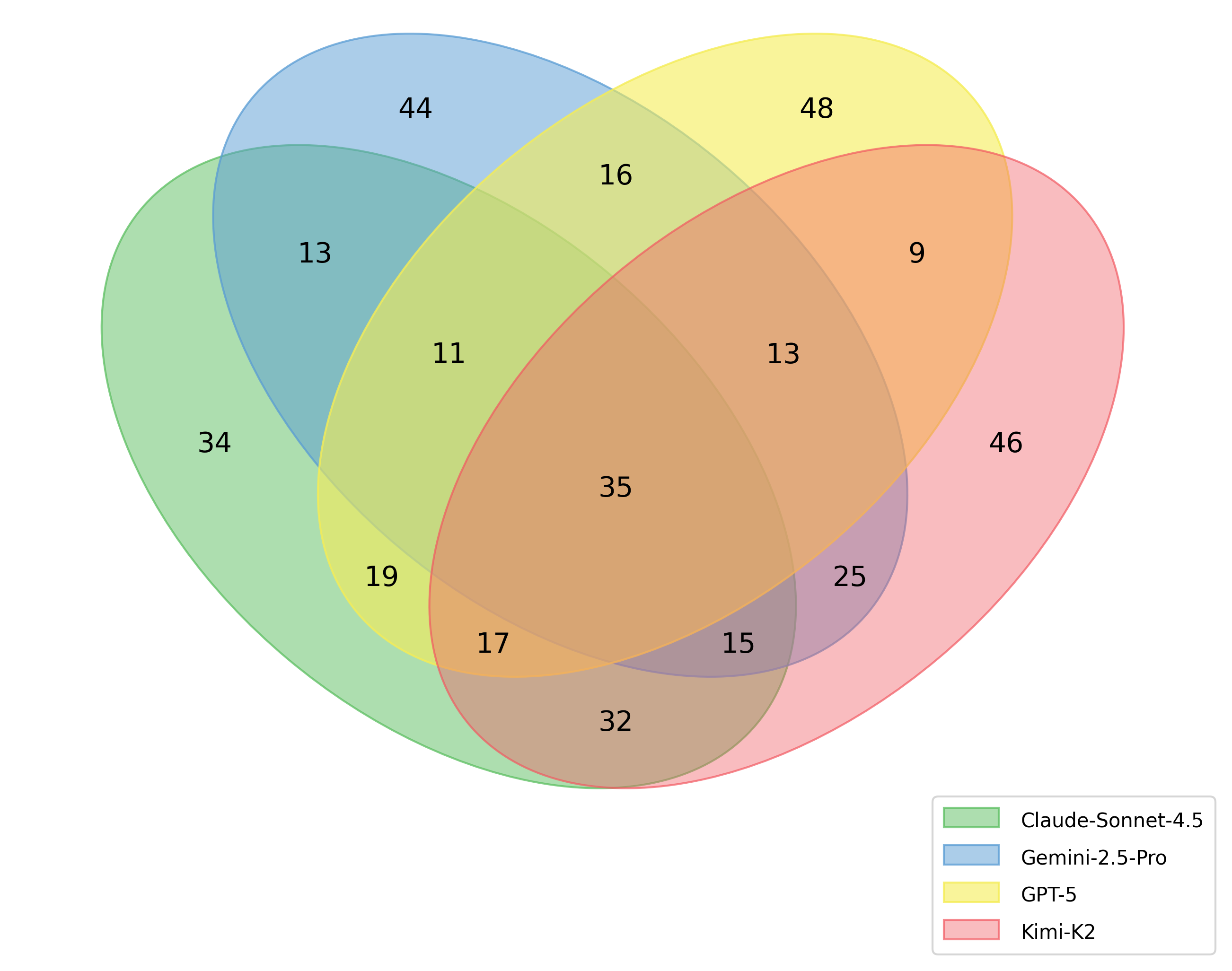}
  \caption{Venn Diagram for Violated Design Constraints of Different LLMs.}
  \label{fig:venn_llms}
\end{figure}

\textit{Overall comparison.}
Table~\ref{tab:rq1_overall_performance} shows clear differences across models.
Claude-Sonnet-4.5 achieves the highest design alignment (DSR=50.20\%) and the lowest violation rate (DVR=37.15\%), indicating comparatively stronger compliance with applicable design constraints.
In contrast, the other models exhibit lower DSR (38.34\%--41.50\%) and higher DVR (39.92\%--45.85\%).
Notably, while pass rates vary substantially across models on \benchPro{} (13.83\%--43.87\%), the corresponding differences in DSR are modest, suggesting that design violations remain common even when functional resolution improves.

\textit{Overlap of violated constraints.}
To probe qualitative differences, we analyze which design constraints are violated by each model.
For each model, we take the union of constraints violated by its generated patches across all \benchPro{} instances, and visualize the intersections using the Venn diagram in Figure~\ref{fig:venn_llms}.
Across the four models, the union contains 377 distinct violated constraints, of which 35 (9.3\%) are violated by all models.
This shared core indicates systematic design challenges that current LLMs consistently miss, likely because such constraints encode repository-specific and context-dependent knowledge not captured by general pretraining. 
Figure~\ref{fig:case_django_13410} illustrates one representative example.
For Django \#13410, the design constraint requires catching only \texttt{BlockingIOError} in \texttt{lock()} and avoiding a broad \texttt{OSError} catch to prevent backward-incompatible behavior.
However, such repository-specific design considerations are consistently missed by current agents: the agent-generated patch catches all \texttt{OSError} exceptions, thereby violating the constraint and potentially introduce security or reliability risks.
Meanwhile, each model exhibits a non-trivial set of uniquely violated constraints.
For example, the largest model-specific set contains 48 constraints (12.7\% of the union), suggesting differences in how models attend to or reason about design-relevant signals during issue resolution.

% \textit{Takeaway.}
% Overall, stronger models reduce the \emph{frequency} of design violations, but design alignment remains far from solved.
% Moreover, the substantial overlap in violated constraints suggests that many failures stem from common limitations of current LLMs, rather than deficiencies of a particular model or agent configuration.

\begin{figure}[t]
\centering
\begin{compactcase}{Case: Django \# 13410}
  \begin{tcolorbox}[
    colback=header_gray,
    boxrule=0pt,
    % frame hidden,
    sharp corners,
    left=4pt,right=4pt,top=2pt,bottom=2pt,
    boxsep=0pt,
    after skip=0pt]
    \fontsize{6pt}{7.5pt}\selectfont
    \textbf{Constraints:}Only catch BlockingIOError in lock(), and do not catch OSError (it's backward incompatible); keep other OSError exceptions propagated.
  \end{tcolorbox}

  \vspace{4pt}

  \noindent\hspace{4pt}%
  \begin{minipage}[t]{0.47\linewidth}
    \vspace{0pt}
    \begin{patchbox}[c2]{patch_gray}{Agent-Generated Patch}
\begin{lstlisting}[style=minicode,aboveskip=0pt,belowskip=0pt,language=Python]
try:
    fcntl.flock(_fd(f), flags)
    return True
except OSError:
    return False
\end{lstlisting}
    \end{patchbox}
  \end{minipage}%
  \hfill
  \begin{minipage}[t]{0.47\linewidth}
    \vspace{0pt}
    \begin{patchbox}[c2]{gold_white}{Gold Patch}
\begin{lstlisting}[style=minicode,aboveskip=0pt,belowskip=0pt,language=Python]
try:
    fcntl.flock(_fd(f), flags)
    return True
except BlockingIOError:
    return False
\end{lstlisting}
    \end{patchbox}
  \end{minipage}\hspace{4pt}

  \vspace{3pt}
\end{compactcase}
  \caption{An example of design-constraint violation in Django \#13410.}
  \label{fig:case_django_13410}
\end{figure}

\summary{
Across four different LLMs, design satisfaction varies modestly and design violations remain prevalent.
Violation overlap reveals both a shared core of systematically missed, system-specific constraints and model-specific blind spots.
}

\subsection{RQ4: Investigation on Design Satisfaction Improvement}

Given the importance of design constraints for long-term software maintainability and the high violation rates exhibited by current LLM-based agents, this research question examines whether explicitly providing design-constraint guidance can reduce design violations.
To this end, agents are supplied with extracted, issue-specific design constraints and instructed to refine their initially generated patches. This setting reflects real-world development, where developers submit an initial patch and iteratively refine it based on reviewer feedback.
We then compare the refined patches with the original ones in terms of \emph{Design Violation Rate (DVR)} and \emph{Pass Rate}.

\begin{figure}[bp]
  \centering
  \includegraphics[width=\linewidth]{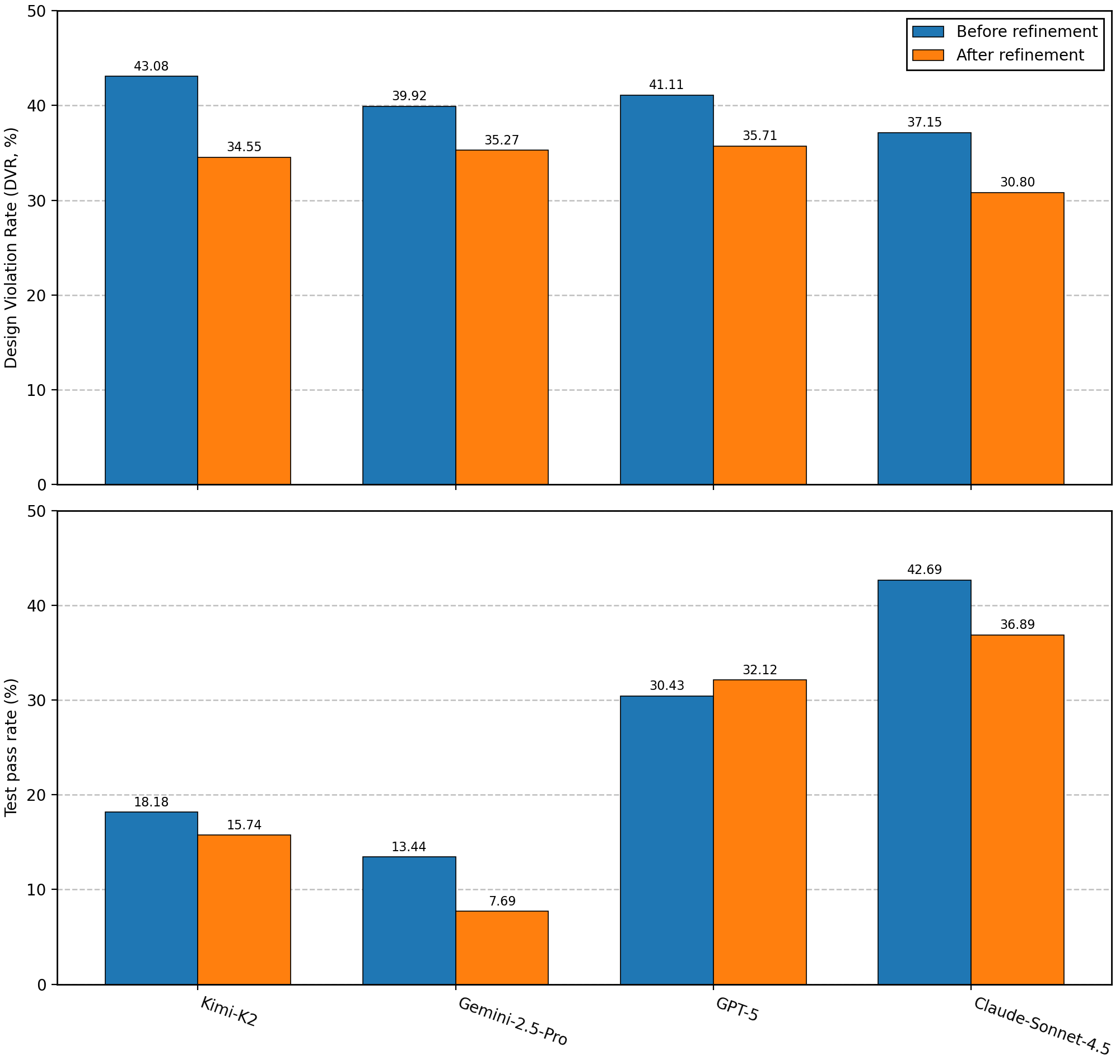}
  \caption{Comparison of DVR Before and After Constraint-Guided Refinement.}
  \label{fig:dvr_before_after}
\end{figure}

\textit{Change in design violations.}
Figure~\ref{fig:dvr_before_after} presents the results.
Across all evaluated agents, incorporating explicit design-constraint guidance leads to a clear and consistent reduction in DVR.
Compared to the original patches, the refined versions violate fewer design constraints, indicating that many design violations stem from missing project-specific design knowledge.
These improvements suggest that making design knowledge explicit is an effective way to mitigate this gap: once relevant design rationales are surfaced, agents are more likely to revise patches toward design-compliant solutions and avoid superficially correct but design-incompatible fixes.
Notably, however, DVR remains above 30\% even after refinement, which may be because current models still struggle to correctly operationalize the provided design constraints. Figure~\ref{fig:case_django_50909} shows a representative example.
Although the agent is explicitly provided with the design constraint during refinement, the refined patch only partially follows the intended guidance: it recognizes that the existing \texttt{min}/\texttt{max} implementations (guarded by \texttt{HAS\_MIN\_MAX}) should be reused, but it fails to prioritize this rule as the primary branch.
As a result, in some cases the patch still falls back to re-implementing logic locally, deviating from the repository-preferred design choice. This highlights that design satisfaction remains far from solved and motivates further research on more effective design-aware reasoning mechanisms.

\textit{Variation across models.} The magnitude of DVR reduction varies across foundation models, suggesting that they absorb and apply the provided design knowledge to different extents. For example, Claude-Sonnet-4.5 reduces DVR from 37.15\% to 30.80\%, whereas Gemini-2.5-Pro decreases it from 39.92\% to 35.27\%.
This variation indicates that, even with identical design guidance, models differ in how effectively they internalize and operationalize design constraints during patch revision.

\textit{Change in Pass Rate.}
Figure~\ref{fig:dvr_before_after}  reports the test pass rates of both the initial and refined patches. The results reveal a trade-off between design compliance and functional correctness during refinement: only GPT-5 achieves a slight improvement, while the other models exhibit regressions relative to their initial pass rates. This indicates that, although providing explicit design constraints can reduce design violations, current LLMs still struggle to enforce these constraints without compromising existing functionality, emphasizing the need for more advanced approaches that can jointly maintain design compliance and functional correctness.

% The magnitude of DVR reduction varies across foundation models, implying that different models absorb and apply the provided design knowledge to different extents. For example, after refinement, Claude-Sonnet-4.5 reduces DVR from 37.15\% to 30.80\%, while Gemini-2.5-Pro reduces DVR from 39.92\% to 35.27\%.

\summary{
Providing explicit, issue-specific design-constraint guidance consistently reduces design violations across models, but residual violation rates remain high.
This suggests that while surfacing design knowledge is a necessary step toward better design satisfaction, more advanced mechanisms are required to fully internalize and reason over design constraints.
}

\begin{figure}[t]
\centering
\begin{compactcase}{Case: Ansible \# 50909}
  \begin{tcolorbox}[colback=header_gray, boxrule=0pt, sharp corners, 
                    left=4pt, right=4pt, top=2pt, bottom=2pt, boxsep=0pt, after skip=0pt]
    \fontsize{6pt}{7.5pt}\selectfont
    \textbf{Constrains:} Import the `min` and `max` filters from jinja2 and use those implementations rather than re-implementing them locally.
  \end{tcolorbox}
  
  \nointerlineskip\vspace{4pt}
  
  \noindent\hspace{4pt}%
  \begin{minipage}[t]{0.47\linewidth}
    \vspace{0pt}
    \begin{patchbox}[c2]{patch_gray}{Agent-Generated Patch after Refinement}
\begin{lstlisting}[style=minicode,language=Python]
def min(environment, a, **kwargs):
    if kwargs:
        if HAS_MIN_MAX:
            return do_min(environment, a, **kwargs)
        else:
        // ...
    else:
        _min = __builtins__.get('min')
        return _min(a)
\end{lstlisting}
    \end{patchbox}
  \end{minipage}%
  \hfill
  \begin{minipage}[t]{0.47\linewidth}
    \vspace{0pt}
    \begin{patchbox}[c2]{gold_white}{Gold Patch}
\begin{lstlisting}[style=minicode,language=Python]
def min(environment, a, **kwargs):
    if HAS_MIN_MAX:
        return do_min(environment, a, **kwargs)
    else:
        if kwargs:
            ...
        _min = __builtins__.get('min')
        return _min(a)
\end{lstlisting}
    \end{patchbox}
  \end{minipage}\hspace{4pt}
  
  \vspace{3pt}
\end{compactcase}
  \caption{An example of design-constraint violation in Django \#50909.}
  \label{fig:case_django_50909}
\end{figure}

\section{Discussion}
In this section, we discuss the limitations and key insights uncovered in our work, highlighting potential directions for future research in design-aware issue resolution and related areas.

% \paragraph{Limited scope of datasets and agents.} \chong{This is not limitation, it is an external threat instead. The limitations should reveal the technical disadvantages of our approach, including constraint extraction and satisfaction verification.}
% The construction of \bench{} currently relies on two datasets, SWE-Bench-Pro and SWE-Bench-Verified.
% While these datasets together cover \xx{} repositories and span diverse real-world projects, they may not fully represent the spectrum of design practices and constraints in other software domains.
% Future work could extend \bench{} to more complex software systems to further improve coverage and strengthen the generality of our findings.
% Moreover, our evaluation considers \xx{} representative agents built upon \xx{} state-of-the-art large language models.
% Although these agents reflect a range of contemporary LLM-based issue-resolution systems, the rapidly evolving model and agent landscape motivates broader evaluation across more foundation models, agent architectures, and tool-use configurations.

% \paragraph{False Negatives in Design Constraint Extraction.}

\textit{Scalability Challenges in Constraint Extraction for Large Codebases.}
Large code repositories often contain extensive histories, including thousands of pull requests, long-running review threads, and evolving design conventions. Extracting design constraints from such repositories can face scalability challenges due to LLM computation cost and the overhead of maintaining traceability across massive and heterogeneous artifacts. Techniques such as hierarchical aggregation, incremental extraction, or selective sampling may be needed to handle enterprise-scale software efficiently.

\textit{Limitations in ``Gold'' Patches Regarding Design Considerations.}
Manual inspection of existing issue resolution datasets revealed that some developer-approved patches satisfy functional tests but do not fully adhere to relevant design constraints. This limitation highlights a key gap in traditional benchmarks, which primarily evaluate correctness through test suites derived from ``gold'' patches. It also motivates future research into assessing and improving the quality of ``gold'' patches. Our patch satisfaction verification method offers a potential tool for addressing this gap.

\textit{Opportunities for Formal Verification of Design Compliance.}
While our benchmark relies on LLM-based verification to assess design satisfaction, formal verification techniques could provide stronger guarantees for certain classes of constraints. For example, static checkers synthesized with LLM support, type-based reasoning, or model checking could complement LLM assessments and enhance reliability, particularly in high-assurance software contexts.

\textit{Requirements for Effective Design-Aware Issue Resolution Techniques.}
Our study of design satisfaction improvements suggests that passing functional tests alone is insufficient for achieving high-quality design alignment. Design-aware issue resolution systems must understand the underlying design rationale, reason about alternative solutions, and account for context-specific applicability conditions. Future approaches should combine semantic reasoning over design constraints with structured knowledge representations to guide patch generation effectively.

% \paragraph{False Negatives in Design Constraint Extraction.}
% Our design satisfaction analysis relies on the extracted design constraints associated with each issue.
% As with any attempt to formalize implicit developer knowledge, not all relevant design considerations are guaranteed to be captured, particularly those that are undocumented, tacit, or only enforced through long-term project culture.
% As a result, a patch labeled as design-aligned under our constraint set may still violate unobserved design expectations that fall outside the benchmark annotations.
% In constructing \bench{}, we prioritize precision in constraint extraction and validation, including only constraints well supported by evidence from project artifacts. Importantly, even under this setting, we still observe consistently low design satisfaction and frequent design violations across agents, underscoring that design adherence remains a critical yet largely unmet requirement for agentic issue resolution.

% \paragraph{Limited 探索 of Design .}

\section{Threats to Validity}

\textit{Internal Validity.}
A primary threat to internal validity arises from the inherent stochasticity of LLM outputs. To mitigate this effect, we fix the decoding temperature to zero for all models.
In addition, our pipeline uses LLMs for both design constraint extraction and design satisfaction judgment, which may threaten internal validity. To alleviate this threat, we conduct manual inspections of extracted constraints and a sampled set of LLM-based judgments to verify correctness and consistency.
For human evaluation, we employ multiple annotators, provide detailed annotation guidelines, and measure inter-annotator agreement to promote consistent judgments and reduce subjective bias.

% A primary threat to internal validity arises from the inherent stochasticity of LLM outputs. To mitigate this effect, we fix the decoding temperature to zero for all models. Since LLMs are also used for both design constraint extraction and satisfaction judgment, we further mitigate potential bias through manual inspection of extracted constraints and a sampled set of model decisions. For human evaluation, multiple annotators, detailed guidelines, and inter-annotator agreement are adopted to reduce subjectivity and promote consistency.

% \paragraph{External Validity.}
% Our evaluation currently focuses on design constrains extracted from pull request discussions.
% However, design decisions are also commonly documented in other artifacts and platforms (e.g., issue trackers), which could affect the external validity of our findings. Given the analogous metadata structures across discussion platforms (e.g., threaded replies and linked code changes), we anticipate that our approach can be adapted to these settings.
% Nonetheless, further evaluation across additional artifacts is warranted to validate this assumption and to better characterize the portability of \bench{}.

\textit{External Validity.}
The external validity of our study mainly stems from the limited scope of our experimental setting. \bench{} is currently constructed on two datasets, SWE-Bench-Pro and SWE-Bench-Verified, covering 2 repositories, and our experiments involve 3 representative agents built upon 3 state-of-the-art foundation models. While these settings span diverse real-world projects and contemporary agentic systems, extending the evaluation to more datasets, repositories, and model/agent variants would provide stronger evidence for generalization.

\textit{Construct Validity.}
Design satisfaction is an inherently abstract concept that cannot be directly observed.
Our metrics operationalize design satisfaction through explicit, instance-level design constraints derived from issue discussions and related artifacts.
While this formulation enables systematic evaluation, it may not fully capture all aspects of design intent, particularly tacit or undocumented decisions.
Thus, DSR and DVR should be interpreted as approximations of design satisfaction rather than exhaustive measures. Still, this limitation does not diminish the utility of the evaluation: compliance with such explicit project-specific, long-tail constraints is a necessary condition for producing high-quality patches, and therefore remains a meaningful indicator of an LLM's design-awareness beyond functional correctness.

% \paragraph{Construct Validity.}
% Design satisfaction is an inherently abstract concept that cannot be directly observed.
% Our metrics operationalize design satisfaction through explicit, instance-level design constraints derived from issue discussions and related artifacts.
% While this formulation enables systematic evaluation, it may not fully capture all aspects of design intent, particularly tacit or undocumented decisions.
% Thus, DSR and DVR should be interpreted as approximations of design satisfaction rather than exhaustive measures.

% \paragraph{External Validity.}
% Our conclusions are based on experiments conducted on two benchmarks and a fixed set of agents.
% Although these benchmarks cover a substantial number of repositories and real-world issues, the observed trends may not generalize to all software projects, programming languages, or development contexts.
% In particular, projects with different review cultures or weaker test suites may exhibit different relationships between functional correctness and design satisfaction.

% \paragraph{Reliability.}
% To enhance reliability, we standardize prompts, evaluation procedures, and hyperparameters across all experiments.
% However, as LLM-based systems evolve, reproducing identical behaviors with future model versions may be challenging.
% We therefore release all prompts, evaluation scripts, and benchmark annotations to facilitate reproducibility and independent verification.

\section{Related Work}

\subsection{Design Knowledge Extraction}

Design knowledge accumulated throughout the software lifecycle is critical to long-term maintainability and evolvability. A growing body of research~\citep{DBLP:conf/icml/GruberBBW91, DBLP:journals/jss/JansenBA08, DBLP:conf/kbse/ShiJYCZMJW21, DBLP:conf/icse/SharmaSS21} has therefore explored automatically mining design knowledge from diverse development artifacts, such as emails and issue discussions, motivated by the observation that many important decisions are made and refined through informal communication. 
Among these approaches, Dhaouadi \emph{et al.} proposed Kantara~\citep{DBLP:conf/kbse/DhaouadiOF22}, which aims to automatically construct rationale and decision graphs from commits and further instantiates the framework with LLM-based extraction~\citep{DBLP:journals/pacmse/DhaouadiOF25}. DRMiner~\citep{DBLP:conf/kbse/ZhaoY0LY024, DBLP:conf/kbse/ZhaoY0LY024a} similarly combines LLMs with heuristic signals and decomposes issue discussions into multiple classification tasks to mine latent design information. Other recent studies, such as Zhou \emph{et al.}~\citep{DBLP:journals/corr/abs-2504-20781}, leverage LLMs to generate design rationales for architectural decisions from textual artifacts. Collectively, these approaches primarily aim to recover decisions and the \emph{rationales} underlying them.

In contrast, \tool{} focuses on extracting \emph{design constraints} that operationalize decision knowledge. Beyond capturing ``why'', \tool{} explicitly models ``when it holds'', which is essential for verifying whether subsequent changes respect established design practices. Moreover, instead of casting extraction as sentence classification, \tool{} mitigates the \emph{tangling and scattering} of design concerns across artifacts by consolidating fragmented signals into a unified, constraint-centric representation. The extracted design knowledge is tightly grounded in code, enabling direct traceability between design reasoning and implementation-level evidence.

\subsection{LLM Benchmarking in Software Engineering}
% 以前的数据集/评估工作

Evaluation of code LLMs and agents has evolved from isolated function-level code generation(e.g., HumanEval~\citep{DBLP:journals/corr/abs-2107-03374}) to substantially more complex, repository-level software engineering tasks~\citep{ding2026octobench, li2026advances}. For example, SWE-bench~\citep{DBLP:conf/iclr/JimenezYWYPPN24} evaluates real-world issue resolution on open-source projects, requiring models to understand project context, modify multiple files, and satisfy existing tests. Subsequent benchmarks further move toward enterprise-like settings, where issue resolution spans longer horizons, involves richer dependencies, and is generally more difficult, as exemplified by SWE-bench Pro~\citep{DBLP:journals/corr/abs-2509-16941} and SWE-Lancer~\citep{DBLP:conf/icml/MiserendinoWPH25}. In addition, recent extensions broaden evaluation coverage along multiple axes, including programming languages and input modalities, such as SWE-bench Multilingual~\citep{khandpur2025swebench} and SWE-bench Multimodal~\citep{DBLP:conf/iclr/YangJZLYWPMSNY025}. Despite improved realism in task setting and issue complexity, the dominant evaluation protocol in these benchmarks still centers on functional correctness, typically operationalized as test-case pass rates, providing an incomplete picture of resolution quality.

Recent work has also begun to refine evaluation objectives for issue resolution beyond functional correctness, introducing metrics such as efficiency and safety~\citep{DBLP:journals/corr/abs-2507-12415,DBLP:journals/corr/abs-2511-05459,DBLP:journals/corr/abs-2511-06090}. However, evaluating LLMs and LLM-based agents in terms of their awareness of design constraints remains insufficiently explored. A related line of work~\citep{ding2026octobench, kottamasu2026apex} attempts to incorporate additional quality signals by using explicit checklists or heuristic criteria, which are either manually curated or generated by LLMs, to assess whether model outputs adhere to specified requirements. These evaluations primarily emphasize instruction following and generic quality attributes. As a result, checklist-based protocols often fail to capture project-specific design decisions that emerge organically through real development processes and are rarely formalized as explicit rules.

In contrast, this work grounds design-aware evaluation in implicitly expressed design constrains mined directly from code review discussions. Rather than relying on externally imposed checklists, the proposed framework derives design constraints from authentic, project-native deliberations and evaluates LLMs and agents on their ability to recognize and comply with these constraints during issue resolution.

% % 近期有部分重合的工作
% Recently, a small number of studies have begun to recognize that evaluating LLM-based code generation and repair solely based on test pass rates is insufficient. These works attempt to incorporate additional quality signals by introducing explicit checklists or heuristic criteria, often constructed manually or generated by LLMs themselves. While such approaches move beyond pure functional correctness, they remain largely detached from the actual design knowledge that governs real software systems.

% In particular, these checklist-based evaluations do not capture project-specific design decisions that emerge organically through development processes, nor do they address how such decisions are implicitly negotiated, justified, and refined in code review discussions. As a result, the evaluated criteria are often generic, externally imposed, and insensitive to repository-specific design practices.

% In contrast, our work grounds design-aware evaluation in *implicitly expressed design decisions* mined directly from code review discussions. We introduce a dedicated design knowledge extraction pipeline that explicitly addresses the inherent mixing and scattering of design knowledge in review comments, reconstructing coherent design decisions from fragmented and entangled evidence. This extracted design knowledge is then used to support design-aware evaluation under realistic contextual trade-offs, rather than relying on predefined or surface-level checklists.

\section{Conclusion}
In this paper, we argue that evaluating LLM-based issue resolution solely through functional correctness provides an incomplete view of patch quality in real-world software. We present \bench, a benchmark for \emph{design-aware issue resolution evaluation} that makes implicit, project-specific design constraints explicit, traceable, and measurable. By extracting constraints from historical pull requests and code review discussions and grounding them in concrete issue resolution tasks, \bench reveals a gap between test-based success and true design alignment. Our results show that state-of-the-art LLM-based agents often produce patches that pass all tests yet violate design constraints, and that functional correctness exhibits little statistical dependence on design compliance. While surfacing relevant design knowledge can reduce violations, many persist, indicating current models struggle to operationalize implicit design rationale. \bench enables systematic study of design alignment, fine-grained diagnosis of design failures, and development of design-aware issue resolution techniques, highlighting the need to move evaluation from \emph{test-passing} toward \emph{design-respecting} as a first-class objective.

% \section{Data Availability}
% Our data and code are included in our replication package~\cite{sweShield}.
% All of the artefacts and LLM output are made publicly available and can be accessed here:\\ 

% \begin{acks}
% \end{acks}

\bibliographystyle{ACM-Reference-Format}
\bibliography{ref}
\balance

\end{document}